\documentclass[12pt]{article}
\usepackage{amsmath,amssymb,natbib,amsbsy,mathrsfs,amsfonts} 
\usepackage{graphicx}
\usepackage{multirow}
\usepackage{tablefootnote}
\usepackage{threeparttable}
\usepackage{pdflscape}
\usepackage{enumerate}
\usepackage{pdfpages}
\usepackage{threeparttable}
\usepackage{comment}
\usepackage{mathrsfs}
\usepackage{lscape}
\usepackage{float}
\usepackage{pdfpages}
\usepackage{authblk}

\usepackage{algorithmicx}
\usepackage[ruled]{algorithm}
\usepackage{algpseudocode}
\usepackage{color}
\usepackage{array}
\newcolumntype{H}{>{\setbox0=\hbox\bgroup}c<{\egroup}@{}}

\DeclareMathOperator{\sign}{sign}
\DeclareMathOperator*{\argmin}{arg\,min}

\newtheorem{theorem}{Theorem}

\def \R { \mathbb{R} }

\newcommand{\T}{{\mbox{\scriptsize \sf T}}}
\newcommand{\dd}{\mathrm{d}}
\newcommand{\Ld}{\Lambda}
\newcommand{\ld}{\lambda}
\newcommand{\af}{\alpha}
\newcommand{\ta}{\theta}

\newcommand{\bbt}{\boldsymbol{\beta}}
\newcommand{\baf}{\boldsymbol{\alpha}}
\newcommand{\bphi}{\boldsymbol{\phi}}
\newcommand{\bpsi}{\boldsymbol{\psi}}
\newcommand{\bta}{\boldsymbol{\theta}}
\newcommand{\bsigma}{\boldsymbol{\sigma}}

\newcommand{\diag}{\text{Diag}}

\newcommand{\bu}{\mathbf{u}}
\newcommand{\bv}{\mathbf{v}}
\newcommand{\bh}{\mathbf{h}}

\newcommand{\bq}{\boldsymbol{q}}
\newcommand{\bld}{\boldsymbol{\lambda}}

\newcommand{\bX}{\mathbf{X}}

\newcommand{\lag}{t}
\newcommand{\blag}{\lag}
\graphicspath{{.}{./Figure/}}

\begin{document}

\title{Regularized estimation for highly multivariate log Gaussian Cox processes}

\author[1]{Achmad Choiruddin}
\author[1]{Francisco Cuevas-Pacheco}
\author[2]{Jean-Fran\c cois Coeurjolly}
\author[1]{Rasmus Waagepetersen}

\affil[1]{Department of Mathematical Sciences, Aalborg University}
\affil[2]{Department of Mathematics, Universit\'e du Qu\'ebec \`a Montr\'eal}


		\maketitle
		
		\begin{abstract}
			Statistical inference for highly multivariate point pattern data is challenging due to complex models with large numbers of parameters. In this paper we develop numerically stable and efficient parameter estimation and model selection algorithms for a class of multivariate log Gaussian Cox processes. The methodology is applied to a highly multivariate point pattern data set from tropical rain forest ecology. 
		\end{abstract}
		

	\section{Introduction}
	Highly multivariate point pattern data are becoming increasingly common. Tropical rain forest ecologists, for example, collect data on locations of thousands of trees belonging to hundreds of species. Likewise, huge space-time data sets regarding scene, time and type of crimes are recorded and made publicly available for many major cities across the world. Research on statistical methodology for multivariate point patterns has mainly considered bivariate or trivariate point patterns. Some exceptions are \cite{diggle:etal:05} and \cite{baddeley:jammalamadaka:nair:14} who considered four- and six-variate multivariate Poisson processes and more recently \cite{jalilian:etal:15} and \cite{waagepetersen:etal:16} who considered five- and nine-variate multivariate Cox processes. A truly high-dimensional analysis was conducted by \cite{rajala:murrell:olhede:17} who introduced a multivariate Gibbs point process and applied it to a point pattern data set containing locations of 83 species of rain forest trees. 
	
	A particular challenge regarding modeling of highly multivariate point patterns is that models easily become very complex with large numbers of parameters. To enhance interpretability of fitted models and numerical stability of estimation, \cite{rajala:murrell:olhede:17} used regularization methods such as the group lasso. The possibility of using regularization was also mentioned in the discussion of \cite{waagepetersen:etal:16} in the context of multivariate log Gaussian Cox processes. 
	
	The type of multivariate log Gaussian Cox process considered
	by \cite{waagepetersen:etal:16} and reviewed in
	Section~\ref{sec:mlgcp} has a simple and natural
	interpretation and e.g.\ enables the user to decompose
	variation according to different sources and to group
	different types of point patterns according to similarities in
	their spatial distributions, see \cite{waagepetersen:etal:16}
	for details. However, the fitting of these models is very
	challenging in the highly multivariate case due to model complexity. In
	Section~\ref{sec:rlse} of this paper, we develop a numerically
	stable and efficient parameter estimation methodology by
	introducing regularization and using efficient convex
	optimization algorithms. We test the methodology in a
	simulation study in Section~\ref{sec:sim} and apply it to a
	tropical rain forest data in
	Section~\ref{sec:app}. Section~\ref{sec:disc} contains some
	concluding remarks.

	\section{Multivariate log Gaussian Cox processes}\label{sec:mlgcp}
	
	A multivariate log Gaussian Cox point process \citep[see][]{moeller:syversveen:waagepetersen:98} is a multivariate point process $\bX=(X_1,\ldots,X_p)$, $p>1$, where each component $X_i$, $i=1,\ldots,p$, is a Cox process driven by a log Gaussian random intensity function $\Ld_i$. Conditionally on the $\Lambda_i$, the $X_i$ are independent Poisson point processes each with intensity function $\Ld_i$. As in \cite{waagepetersen:etal:16}, we assume that the random intensity functions are of the form
	$\Ld_i(\bu)=\exp[Z_i(\bu)]$ with
	\begin{align}
	Z_i(\bu)=\mu_i(\bu)+Y_i(\bu)+U_i(\bu), \; \bu \in \R^2. \label{eq:Z}
	\end{align}
	The terms $\mu_i$ are deterministic and typically given in terms of regressions on observed covariates. 
	The terms $Y_i$ and $U_i$ are zero-mean Gaussian fields. The $Y_i$ can be mutually correlated while the $U_i$ are assumed to be independent. The $U_i$ are assumed to be stationary with variances $\sigma_i^2 > 0$ and correlation functions $c_i$, $i=1,\ldots,p$.
	For the $Y_i$ we assume that 
	\[ Y_i(u)= \sum_{l=1}^q \af_{il}E_l(u) \]
	where $q \ge 1$, $\baf=[\af_{ij}]_{ij}$ is a $p\times q$ real valued coefficient matrix, and the $E_l$, $l=1,\ldots,q$, are independent zero-mean stationary Gaussian fields with variance one. In our applications we also consider the case $q=0$ meaning that the $Y_i$ are omitted in \eqref{eq:Z}. The $Y_i$ can be interpreted as effects of unobserved spatial covariates while the $U_i$ represent sources of clustering which are specific to each type of points.  We denote by $r_l$ the correlation function of $E_l$.  
	For the correlation functions $r_l$ and
	$c_i$ we introduce isotropic parametric models $r_l(\cdot;\phi_l)=r(\|\cdot\|/\phi_l)$ and
	$c_i(\cdot;\psi_i)=c(\|\cdot\|/\psi_i)$, where $\phi_l$ and $\psi_i$ are correlation scale parameters.  Specifically, we consider in this paper exponential correlation functions $r(t)=c(t)=\exp(-t)$, $t \ge 0$, although many other choices are available \citep{chiles:delfiner:99}.
	
	\subsection{Intensity function and pair correlation function}\label{sec:moments}
	
    Let $\baf_{i \cdot}$ denote the $i$th row of $\baf$. Following \cite{moeller:syversveen:waagepetersen:98}, the intensity function of $X_i$ is  $\rho_i(\bu) = \exp\big [\mu_i(\bu)   + \baf_{i\cdot} \baf^{\T}_{i\cdot}/2 +\sigma^2_i/2\big ]$ while the cross pair correlation function for the pair $X_i$ and $X_j$ is 
	\begin{equation}\label{eq:crosspair}
	g_{ij}(\blag) = \exp\big[  \sum_{l=1}^{q}\af_{il}\af_{jl}r_{l}(\blag;\phi_l) + 1(i=j)\sigma^2_ic_i(\blag;\psi_i) \big ] \end{equation}
	for $\blag\ge 0$. Consider two spatial locations $\bu$ and $\bv$. Then $\rho_j(\bv)g_{ij}(\|\bv-\bu\|)$ represents the cross-Palm intensity function \citep{coeurjolly:moeller:waagepetersen:17} and can be interpreted as the intensity function of $X_j$ conditional on that $\bu \in X_i$. Hence $g_{ij}(\|\bv-\bu\|)>1$ ($<1$) implies that presence of a point from $X_i$ at $\bu$ increases (decreases) the intensity of $X_j$ at $\bv$. Thus $\sum_{l=1}^{q}\af_{il}\af_{jl}r_{l}(\blag) < 0$ ($>0$) implies repulsion (attraction) between points of $X_i$ and $X_j$ at lag $\blag$. Similarly, a large value of $\sum_{l=1}^{q} \af_{il}^2 r_{l}(\blag)+\sigma^2_ic_i(\blag)$ leads to strong attraction among points of $X_i$ separated by a lag $\blag$.
	
	Non-parametric kernel estimates of the $g_{ij}$ are given by 
	\begin{equation}\label{eq:ghat}
	\hat{g}_{ij}(t) = \frac{1}{2\pi t} \sum_{\substack{\bu \in X_{i} \cap W,\\ \bv \in X_{j} \cap W,\\ \bu \neq \bv}} \frac{k_{b}(t-\|\bu-\bv\|)}{\hat{\rho}_i(\bu)\hat{\rho}_j(\bv) |W \cap W_{\bu- \bv}|}, \quad t>0,
	\end{equation}
	where $W$ is the observation window, $k_{b}$ is a kernel function depending on a smoothing parameter $b>0$, $|\cdot|$ denotes area and $W_{\bh}$ denotes the translate of $W$ by the vector $\bh \in \R^{2}$ \citep{moeller:waagepetersen:03}. The quantities $\hat{\rho}_i$ and $\hat{\rho}_j$ are estimates of the intensity functions of $X_i$ and $X_j$, typically obtained from regression models depending on observed covariates through maximizing the composite likelihood \cite[see e.g.][]{waagepetersen:07,moeller:waagepetersen:07} or its regularized versions \cite[e.g.][]{thurman2015regularized,choiruddin2018convex}. 
	
	\subsection{Least squares estimation}\label{sec:lse}
Let  $\bta$ be the parameter vector consisting of the components of
    $\baf$,     $\boldsymbol{\sigma}^2 = (\sigma_{1}^{2},\ldots,\sigma_{p}^{2})^{\T}$, $\boldsymbol{\phi} = (\phi_{1},\ldots,\phi_{q})^{\T},$  and $\boldsymbol{\psi} = (\psi_{1},\ldots,\psi_{p})^{\T}$. Let further 
\begin{align} \bbt_{ij}(\baf,\bsigma^2)& =
                                         (\af_{i1}\af_{j1},\ldots,\af_{iq}\af_{jq})^\T,
                                         i\neq j, \nonumber \\
  \bbt_{ii}(\baf,\bsigma^2) & =
                              (\af^2_{i1},\ldots,\af^2_{iq},\sigma_i^2)^\T. \label{eq:betaii}
\end{align}
	The objective function used by \cite{waagepetersen:etal:16} for parameter estimation is of the form 
	\begin{align}\label{eq:lso}
	Q(\bta) = \sum_{i,j=1}^{p} \| Y_{ij}- X_{ij}(\bphi,\bpsi) \bbt_{ij}(\baf,\bsigma^2) \|^2,
	\end{align}
	where
	\[ Y_{ij}= (\sqrt{w_{ij1}}\log \hat{g}_{ij}(t_1),\ldots, \sqrt{w_{ijL}}\log \hat g_{ij}(t_L))^\T, \]
	$\hat g_{ij}(t_k)$, $k=1,\ldots,L$,  are obtained using \eqref{eq:ghat}
	for lags $0<t_1 < t_2 < \ldots <t_L$ and the $w_{ij} \ge 0$ are non-negative weights.  The matrix $X_{ij}(\bphi,\bpsi)$ is $L \times q$ ($i\neq j$) or $L \times (q+1)$ ($i=j$) with rows $\sqrt{w_{ijk}}\mathbf r(t_k;\bphi)$ ($i \neq j$) or $\sqrt{w_{iik}} [ \mathbf r(t_k;\bphi),c_i(t_k;\psi_i)]$ ($i=j$), $k=1,\ldots,L$,
	where  
	\begin{align*} 
	\mathbf r(t_k;\bphi)= (r_1(t_k;\phi_1),\ldots,r_q(t_k;\phi_q)). 
	\end{align*}
	\cite{waagepetersen:etal:16} minimized $Q(\bta)$ using a standard quasi-Newton method.
	
	\subsection{Inference regarding multivariate dependence structure}\label{sec:multvarinference}
	The model \eqref{eq:Z} enables us to decompose the covariances of the latent Gaussian fields $Z_i$ into contributions from the common fields $E_l$ and the type-specific fields $U_i$. Specifically, \cite{waagepetersen:etal:16} considered for each type $i$  and lag $t$ the proportion of variance (PV) due to the common fields:
		\begin{align*}
		\mathrm{PV}_i(t) &=\frac{\mathrm{cov}\{Y_i(\bu), Y_i(\bu + \bh)\}}{\mathrm{cov}\{Z_i(\bu), Z_i(\bu + \bh)\}} \\ &=\frac{\sum_{l=1}^{q}\af_{il}^2r_{l}(\blag;\phi_l)}{\sum_{l=1}^{q}\af_{il}^2r_{l}(\blag;\phi_l) + \sigma^2_ic_i(\blag;\psi_i)} , \quad \|\bh\|=t.
		\end{align*}
		These are useful e.g.\ for grouping species based on how much of the variation is due to common factors respectively type-specific factors.  Furthermore,  from $\baf$ and $\bsigma^2$ we can compute the matrix of lag zero inter-type covariances $\baf \baf^\T$ due to the common latent fields with $ij$th entry 
\[ \mathrm{cov}\{Y_i(\bu), Y_j(\bu)\}=\baf_{i.}\baf_{j.}^\T 
\] 
as well as the lag zero covariances between the fields including both common and type-specific effects,
\begin{equation}\label{eq:covZ} \mathrm{cov}\{Z_i(\bu), Z_j(\bu)\}=\baf_{i.}\baf_{j.}^\T +1[i=j]\sigma^2_i. 
\end{equation}
A row $\baf_{i\cdot}$ informs on the dependence of $X_i$ on the common latent fields. Considering the norms of 
differences $\|\baf_{i.}- \baf_{j.} \|$, we are able to group the
different types of point patterns according to their dependence on the
latent factors $E_l$. 

 As discussed in \cite{waagepetersen:etal:16}, the distribution of our multivariate log Gaussian Cox process is invariant to 1) simultaneous permutation of columns in $\baf$ and corresponding $\phi_i$'s and 2) multiplication of a column in $\baf$ by $-1$. Thus we can not identify individual parameters $\af_{il}$ and $\phi_l$ without imposing constraints on the parameter space. 

In our simulation studies in Section~\ref{sec:sim}, we therefore follow
\cite{waagepetersen:etal:16} by restricting attention to identifiable
functions of $\baf$ and $\boldsymbol{\psi}$ such as the aforementioned
proportions of variances and covariances and norms of differences between rows of $\baf$. In the application, we also consider the percentage of zero entries when $\baf$ is estimated using elastic net regularization with $\xi>0$, see next section. The more zeros, the less complex is the dependence structure of the multivariate log Gaussian Cox process.
	
	\section{Regularized least squares estimation}\label{sec:rlse}
	
	The parameter vector $\bta$ is of potentially very high
        dimension, especially due to the many components of the $p\times q$ parameter matrix $\baf$. To enhance interpretability and numerical stability of estimation we suggest to introduce regularization and thus consider the regularized least squares criterion
	\begin{align}\label{eq:reglso}
	Q_\ld(\bta) = \; Q(\bta)+ \ld \sum_{i=1}^p\sum_{l=1}^q p(\af_{il})
	\end{align}
	where $Q(\bta)$ is given by \eqref{eq:lso}, $\ld$ is a nonnegative tuning parameter and $p(\cdot)$ is a convex penalty function. We consider in the following the elastic net penalization \citep{zou:hastie:05} $p(\af_{il})=(1-\xi) \af_{il}^2/2+\xi |\af_{il}|$, $0 \le \xi \le 1$, which  embraces LASSO \citep{tibshirani:96} and ridge regression \citep{hoerl1988ridge} techniques by setting $\xi=1$ or $\xi=0$ respectively.
	
	Using regularization in a related factor analysis was previously suggested by \cite{choi2010penalized}. Their simpler setting corresponds to directly observing vectors $(Z_i(u_k))_{i=1}^p$, $k=1,\ldots,n$, where $Z_i(u_k)$ is modeled as in \eqref{eq:Z} but with zero spatial correlation. In contrast, our $Z_i$ are unobserved  with spatial correlation modeled via the correlation functions $r_l$ and $c_i$. Thus the computational methodology suggested by \cite{choi2010penalized}  is not applicable in our situation.
	
	To minimize \eqref{eq:reglso} with respect to $\bta$, we  employ a cyclical block descent algorithm where $\bsigma^2$, $\baf$, $\bphi$ and $\bpsi$ are updated in turn. The updating is iterated until relative function convergence of the criterion \eqref{eq:reglso}. The details of the block updates are given in the following two sections and Appendices~\ref{prox}-\ref{sec:CDA}. Pseudo-code for the full algorithm is given in Appendix~\ref{algorithm}.
	
	\subsection{Update for $\bsigma^2$ and $\baf$} \label{sec:afsigma}
	
	Our strategy for updating $\bsigma^2$ and $\baf$ is to use for $i=1,\ldots,p$, a least squares update of
	$\sigma^2_i$ followed by an update of $\baf_{i\cdot}$ using a cyclical
	coordinate descent algorithm. The motivation for updating rows
        $\baf_{i\cdot}$  instead of other subsets of $\baf$ is
        that the update of $\baf_{i \cdot}$, keeping all other
        parameters fixed, is quite close to a standard least squares
        problem, as will be evident in the following.

 The relevant part of the objective
	function for the updates of $\sigma_i^2$ and $\baf_{i \cdot}$ given all other parameters  is
	\begin{align} \label{eq:blockdescent}
	Q_{\ld,i}(\baf_{i \cdot},\sigma^2_i) =  \; 2 \sum_{\substack{j=1\\j \neq i}}^p \| Y_{ij}-
	\tilde X_{ij}	\baf_{i \cdot} \|^2 +  \|Y_{ii}- X_{ii} \bbt_{ii}(\baf,\bsigma^2)  \|^2+ \ld \sum_{l=1}^q  p(\af_{il})
	\end{align}
	where the $l$th column of $\tilde X_{ij}$ is 
	the $l$th column of $X_{ij}$ multiplied by $\af_{jl}$. In other words, for $i \neq j$, 
	$\tilde X_{ij}= X_{ij} \diag (\af_{j1},\ldots,\af_{jq} )$
        where $\diag (\af_{j1},\ldots,\af_{jq} )$ is the diagonal
        matrix with diagonal entries $\af_{j1},\ldots,\af_{jq}$.  For
        ease of notation we here omit the dependence of  $\tilde
        X_{ij}$ and $X_{ii}$ on the fixed parameters $\bpsi$ and
        $\bphi$. Note that \eqref{eq:blockdescent} is
          equivalent to a standard least squares objective function
          for $\baf_{i \cdot}$ except for the middle term that depends
          on $\af_{il}^2$, $l=1,\ldots,q$, cf.\ \eqref{eq:betaii}.
	
	The minimization of $Q_{\ld,i}$ with respect to $\sigma^2_i$
        only involves the middle term in
          \eqref{eq:blockdescent}. This is a standard least squares problem except that we require $\sigma^2_i$ to be non-negative. Thus, 
	\begin{align*}
	\hat \sigma^2_i & = \max \{ 0,\arg\min_{\sigma^2_i}  Q_{\ld,i}(\baf_{i \cdot},\sigma^2_i) \}.
	\end{align*}
An explicit formula for this update is given in Appendix~\ref{sigma}.

	To update $\baf_{i \cdot}$ (given $\sigma_i^2$ and
	all other parameters), we use a so-called proximal Newton update \cite[][and Appendix~\ref{prox}]{lee:sun:saunders:14} where the middle term in \eqref{eq:blockdescent} is replaced by a quadratic approximation around the current value $\baf_{i \cdot}^{(k)}$. We denote by $\hat Q_{\ld,i}(\baf_{i \cdot},\sigma^2_i|\baf_{i \cdot}^{(k)})$  the resulting approximate objective function (to be detailed in the next paragraph). 
	Since $\hat Q_{\ld,i}(\baf_{i \cdot},\sigma^2_i|\baf_{i \cdot}^{(k)})$ is a regularized linear least squares objective function, minimization can be performed using a standard coordinate descent algorithm \citep[see e.g.][]{hastie:tibshirani:wainwright:15}. 

A very simple quadratic approximation of the middle term of \eqref{eq:blockdescent} is 
	\[ \|Y_{ii}- X_{ii} \bbt_{ii}(\baf,\bsigma^2)  \|^2 \approx \|Y_{ii}- \tilde X_{ii}^k [\baf_{i \cdot}^\T,\sigma_i^2]^\T \|^2,  \]
	where $\tilde X_{ii}^k = X_{ii}\diag\big\{\af_{i1}^{(k)},\ldots,\af_{iq}^{(k)},1\big \}$. Nevertheless, the curvature of this quadratic approximation does not match the curvature of the original term at $\baf_{i \cdot}^{(k)}$. Instead we use a second-order Taylor approximation as detailed in the Appendix~\ref{taylor} which results in the explicit expression for $\hat Q_{\ld,i}(\baf_{i \cdot},\sigma^2_i|\baf_{i \cdot}^{(k)})$ given by
	\begin{align}
	Q_{\ld,i}(\baf_{i \cdot},\sigma^2_i) &  \approx \hat Q_{\ld,i}(\baf_{i \cdot}|\baf_{i \cdot}^{(k)}) \nonumber \\
	& = \sum_{\substack{j=1}}^p \| Y^*_{ij}-
	X^*_{ij}\baf_{i \cdot} \|^2 + \ld \sum_{l=1}^q  p(\af_{il}) \label{finalobj},
	\end{align}
	where \begin{align}
	Y^*_{ij} & = \sqrt{2}Y_{ij}, \mbox{ for } i \neq j, \nonumber \\ 
	X^*_{ij} & = \sqrt{2}X_{ij}D(\af_{j \cdot}^{(k)}), \mbox{ for } i \neq j, \nonumber \\
	Y^*_{ii} & = Y_{ii}+ X_{ii,\cdot (1:q)}\baf_{i \cdot}^{2,(k)} - X_{ii,\cdot (q+1)}\sigma^{2}_i, \nonumber \\ 
	X^*_{ii} & = 2X_{ii,\cdot (1:q)} D(\baf_{i \cdot}^{(k)}) \label{eq:YstarXstar}
	\end{align}
and $X_{ii,\cdot (1:q)}$ denotes the first $q$ columns in $X_{ii}$.

	We obtain
	\begin{align*}
	\hat \baf_{i \cdot} & = \arg\min_{\baf_{i \cdot}}  \hat Q_{\ld,i}(\baf_{i \cdot}|\baf_{i \cdot}^{(k)})
	\end{align*}
	using coordinate descent with an explicit formula for the updates given in Appendix~\ref{alpha}.
	Further, define for some $t>0$,
	\begin{align} \label{direction}
	\baf_{i \cdot}^{(k+1)} = \baf_{i \cdot}^{(k)}+ t (\hat \baf_{i \cdot}-\baf_{i \cdot}^{(k)} ).
	\end{align}
	Thus, $\baf_{i \cdot}^{(k+1)}$ is obtained using $(\hat \baf_{i
		\cdot}-\baf_{i \cdot}^{(k)})$ as a search direction with step size
	controlled by $t$. Following
	\citet[][Proposition~2.3]{lee:sun:saunders:14}, one can show (see Appendix~\ref{theory})
	that $Q_{i,\ld}(\baf_{i \cdot}^{(k+1)})< Q_{i,\ld}(\baf_{i
		\cdot}^{(k)})$ if $t$ is small enough. That is, if the minimization
		of $ \hat Q_{i,\ld}$ is combined with a line search the resulting
		update is guaranteed to decrease the objective function
		$Q_{i,\ld}$ written in \eqref{eq:blockdescent}.

	\subsection{Update for $\bpsi$ and $\bphi$} \label{profap}
	To update $\bphi$ and $\bpsi$ given all other parameters, we first reparameterize the objective function in terms of ${\bf f}=(\log \phi_1,\ldots,\log \phi_q )^\T$ and ${\bf s}=(\log \psi_1,\ldots,\log \psi_p)^\T$. We then update ${\bf f}$ and ${\bf s}$ in turn using a standard quasi-Newton update as implemented in the \texttt{optim} routine in the \texttt{R} language with method \texttt{bfgs} (Broyden-Fletcher-Goldfarb-Shanno update). Finally, we transform back using the exponential to get updates of $\bphi$ and $\bpsi$.

	We also tried other options: joint update of $(\bphi,\bpsi)$
        without log-transformation but introducing box constraints to
        avoid negative values and joint quasi-Newton update of the
        log-transformed parameters $(\bf f, \bf s)$. For simulated data examples, the option with separate updates of $\bf f$ and $\bf s$ performed best.
	
	\subsection{Initialization}
        
       We initialize the components $\baf$ by a sample of independent random normals with mean zero and standard deviation 0.05 while we choose 1 for the initial values of the components in $\bsigma^2$. For $\bphi$ and $\bpsi$ we choose initial values that depend on the scale of the observation window to avoid that the corresponding covariance functions become essentially constant equal to zero (too small initial values) or to one (too large initial values). For the unit square observation window, for example, the initial values for $\bphi$ and $\bpsi$ were chosen randomly from the uniform distribution on $[0.01,0.05]$. Regarding the choice of weights $w_{ijk}$ introduced in Section~\ref{sec:lse}, we follow arguments by \cite{waagepetersen:etal:16} and fix, for $i,j=1,\ldots,p$ and $k=1,\ldots,L$, $w_{ijk}=\hat g_{ij}(t_k)/2$ for $i \neq j$ and $w_{iik}=\hat g_{ii}(t_k)$.
	
	\subsection{Strategy to determine $q$ and regularization parameters $\lambda$ and $\xi$} \label{sec:cv}
        In our applications we consider just a few values $\xi=0$ (ridge), $\xi=0.5$ (mix of ridge and LASSO, i.e. elastic net) and $\xi=1$ (LASSO). For each of the values of $\xi$ we use a two-dimensional $K$-fold cross
	validation approach to select optimal values $\ld_{\text{opt}}$ and $q_{\text{opt}}$ among prespecified values $\ld_1,\ldots,\ld_M$ and $q_1,\ldots,q_N$ \citep[e.g.][Chapter 7]{hastie:tibshirani:friedman:13}. The procedure is as follows.
	\begin{enumerate}[Step 1.]
		\item We split indices $ijk$ ($i,j=1,\ldots,p$ and
		$k=1,\ldots,L$) into $K$ sets $S_1,\ldots,S_K$ (see
		details below). \label{step1}
		\item For each $\ld \in \{\ld_1,\ldots,\ld_M\}$ and $q
		\in \{q_1,\ldots,q_N\}$, we obtain an estimate
		$\boldsymbol{\hat \theta}_c$ by minimizing
		equation~\eqref{eq:reglso} with $w_{ijk}$ replaced
		by 0 for $ijk \in S_c,c=1,\ldots,K$. The cross
		validation score for $\lambda$ and $q$ is then obtained by
		\begin{align}
		\mathrm{CV}(\ld,q)=\frac{1}{K} \sum_{c=1}^{K} \mathrm{CV}_c, \label{eq:CV}
		\end{align}
		where $\mathrm{CV}_c= \sum_{ijk \in S_c} (Y_{ijk}- \hat{Y}_{ijk}(\boldsymbol{\hat \theta}_c))^2$ and $\hat{Y}_{ij}(\boldsymbol{\hat \theta}_c)=X_{ij}(\hat \bphi_c,,\hat \bpsi_c) \bbt_{ij}(\hat \baf_c,\hat \bsigma^2_c)$. \label{step2}
		\item To obtain $\ld_{\text{opt}}$ and $q_{\text{opt}}$, we minimize $\mathrm{CV}(\ld,q)$ w.r.t $\ld$ and $q$, i.e.,
		\begin{align}
		(\ld_{\text{opt}}, q_{\text{opt}})=\argmin_{m=1,\ldots,M,n=1,\ldots,N} \mathrm{CV}(\ld_m,q_n). \label{eq:qopt}
		\end{align}
	\end{enumerate}
	The sets $S_c$  in Step~\ref{step1} need to be chosen
	carefully. First, since $\log(\hat g_{ijk})$ and $\log(\hat
	g_{ijk'})$ are strongly correlated when $k$ and $k'$ are
	close, we leave out blocks of consecutive indices. Second, we
	do not include diagonal indices $iik$ in the sets $S_c$ since
	values $Y_{iik}$ include contributions from the type-specific
	random fields. The diagonal values thus do not provide so much
		information about $q$ and omission of these values further
		makes  the estimation procedure less stable regarding $\bsigma^2$ and $\bpsi$. So, to determine each subset $S_c$, we arrange the $ijk$ with $i<j$ lexicographically in a vector $(121,122,\ldots)$ and split this vector into consecutive blocks of length $b$. These blocks are then assigned to the different $S_c$ at random.
	
	The one standard error (1-SE) rule is an
		alternative way to select $\lambda$ and $q$ based on the CV
		scores obtained from \eqref{eq:CV} \citep[e.g.][]{hastie:tibshirani:friedman:13}. In case of $q$ fixed, the 1-SE rule chooses the largest $\lambda$ for which the CV score is less than the smallest CV score plus one standard deviation. In the case where both $\lambda$ and $q$ is to be selected, we adapt the 1-SE rule by starting with $(\ld_{\text{opt}}, q_{\text{opt}})$ given by \eqref{eq:qopt} and then choosing $(\ld,q)$ to be the smallest $q$ and largest $\lambda$ possible such that the following condition holds:
	\begin{align*}
	\mathrm{CV}(\ld,q) \leq \mathrm{CV}(\ld_{\text{opt}},q_{\text{opt}}) + \mathrm{SE}(\ld_{\text{opt}},q_{\text{opt}}),
	\end{align*}
	where \begin{align*}
	\mathrm{SE}(\ld_{\text{opt}},q_{\text{opt}})=\sqrt{\frac{\sum_{c=1}^{K}(\mathrm{CV}_c-\mathrm{CV}(\ld,q))^2}{(K-1)K}}.
	\end{align*}
	Hence, the 1-SE rule attempts to select the most simple model whose CV score is within one standard error of the minimal  CV score.

Finally, note that when $\xi=0.5$ or $\xi=1$ and $\lambda>0$ is chosen, the resulting estimate of $\baf$ may contain columns that consist entirely of zeros. The effective number $q_{\text{eff}}$ of columns in $\baf$ then becomes smaller than $q_{\text{opt}}$.
	
	\section{Simulation study} \label{sec:sim}
	We conduct two simulation studies to evaluate the regularized least squares technique for parameter estimation and the cross-validation (CV) method to select $q$ and $\lambda$. The setting of the first study corresponds to the simulation study  in \cite{waagepetersen:etal:16}. We first compare the estimates obtained using the new cyclical block descent (CBD) algorithm developed in Section~\ref{sec:rlse} with the method proposed by \cite{waagepetersen:etal:16}. In this regard, we consider values of $q=1,\ldots,5$ and for comparison purposes, we fix $\lambda=0$ since regularization was not used in \cite{waagepetersen:etal:16}. Next we consider only the new algorithm with the objective of comparing different CV options for selecting $q$ and $\lambda$, cf.\ Section~\ref{sec:cv}, and to study the effect of regularization. The second study has the same objective but with a more complex setting for the simulations. In both simulation studies we use $K=8$ for the CV and we only consider the LASSO option ($\xi=1$) for regularization.

To asses the parameter estimates, we consider the root mean squared errors (RMSEs) of the estimates. For a real parameter $\omega$ and estimate $\hat \omega$, the RMSE is	
\begin{align*}
	\mathrm{RMSE}(\hat \omega)=\sqrt{\mathbb{E} \big( (\hat \omega - \omega)^2\big)}.
	\end{align*}
For each of the parameter matrices/vectors $\baf \baf^{\T}$, $\bsigma^2$, $\bpsi$, or the vector of proportions of variances at lag 0 (PV), we evaluate the average of RMSEs for the components in these quantities. For example, we compute the average of RMSEs for each entry in the $p \times p$ matrix $\baf \baf^\T$.

	\subsection{Comparison of methods for least squares estimation} \label{sec:simp5:study}
	The first study follows the one in \cite{waagepetersen:etal:16} for which 200 point patterns in $W=[0,1]^2$ are generated from multivariate log Gaussian Cox processes as defined in Section~\ref{sec:mlgcp}, with $p=5$ and $q=2$. The true parameters are: $\boldsymbol{\sigma}^2=(1,1,1,1,1), \; \boldsymbol{\psi}=(0.01,0.02,0.02,0.03,0.04), \; \boldsymbol{\phi}=(0.02,0.1)$ and
	\begin{align*}
	\boldsymbol\af^\T & =
	\begin{bmatrix}
	\sqrt{0.5} & 1& -1& 0& 0 \\ 
	0& 0& 1& -1& 0.5 \\ 
	\end{bmatrix}.
	\end{align*}
	The trend models $\mu_i(u)=m_i$ are set such that the expected number of points is 1000 for each $i=1,\ldots,5$. A uniform kernel with bandwidth 0.005 is used for the non-parametric estimation of the cross pair correlation function at $L=25$ equispaced lags between 0.025 and 0.25.

For each simulation we compare two methods for minimizing \eqref{eq:reglso} with  $\lambda=0$ and $q \in \{1,\cdots,5\}$:
	\begin{enumerate}
		\item The standard quasi-newton (SQN) optimization algorithm considered by \linebreak \cite{waagepetersen:etal:16} and implemented in the \texttt{R} package \texttt{optimx}. This algorithm updates all parameters jointly.
		\item The new CBD algorithm described in Section~\ref{sec:rlse}.
	\end{enumerate}
The comparison is in terms of minimization of the objective function, computing time and RMSEs. 
		
	
	
	\begin{table}[ht]
		\centering
		\renewcommand{\arraystretch}{1.35}
		\setlength{\tabcolsep}{7pt}
		\caption{Averages of the minimized objective function $Q(\bta)$ given by \eqref{eq:lso} and the computing time (in seconds) based on 200 simulations from a multivariate  log Gaussian Cox process ($p=5, q=2$), modeled with  $q \in \{1,2,3,4,5\}$, for two optimization methods.}
		\begin{tabular}{lrrrrr}
			\hline
			\multirow{2}{*}{Method} & \multicolumn{5}{c}{$q$} \\
			\cline{2-6}
			& 1 & 2 & 3 & 4 & 5 \\ 
			\hline
			& \multicolumn{5}{c}{Minimized objective function} \\
			\hline
			SQN & 6.61 & 4.76 & 5.39 & 6.32 & 4.51 \\
			CBD & 3.55 & 1.96 & 1.73 & 1.62 & 1.57 \\ 
			\hline
			& \multicolumn{5}{c}{Timings (seconds)} \\
			\hline 
			SQN & 0.96 & 1.98 & 3.97 & 6.45 & 8.99 \\
			CBD & 1.99 & 3.11 & 4.26 & 5.30 & 5.92 \\ 
			\hline
		\end{tabular}
		\label{tab:QT}
	\end{table}
	
	\begin{table}
		\centering
		\renewcommand{\arraystretch}{1.15}
		\setlength{\tabcolsep}{3pt}
		\caption{Average RMSEs for $\hat \baf \hat \baf^\T, \hat \bsigma^2$, and $\hat \bpsi$ (see explanation in text) obtained from 200 simulations from a multivariate log Gaussian Cox process ($p=5, q=2$), modeled with $q \in \{1,2,3,4,5\}$. The estimates are obtained by minimizing \eqref{eq:lso} with two optimization methods. Last column shows the percentages of outlying parameter estimates removed in the RMSE calculation.}
		\begin{tabular}{crrrrrr}
			\hline
			\multirow{2}{*}{Method} & \multicolumn{5}{c}{$q$} & \multirow{2}{*}{Outliers (\%)} \\
			\cline{2-6}
			& 1 & 2 & 3 & 4 & 5 &  \\ 
			\hline
			\hline
			\multicolumn{7}{c}{$\hat \baf \hat \baf^\T$} \\
			\hline
			SQN & 0.41 & 0.93 & 1.10 & 1.17 & 1.09 & 10.3 \\ 
			CBD & 0.41 & 0.25 & 0.29 & 0.32 & 0.39 & 0 \\ 
			\hline
			\multicolumn{7}{c}{$\hat {\boldsymbol \sigma^2}$} \\
			\hline
			SQN & 0.58 & 0.54 & 0.44 & 0.89 & 0.98 & 1.1 \\
			CBD & 0.34 & 0.18 & 0.28 & 0.39 & 0.50 & 0 \\ 
			\hline
			\multicolumn{7}{c}{$\hat \bpsi$} \\
			\hline
			SQN & 0.0791 & 0.1752 & 0.1337 & 0.4091 & 0.4566 & 11.5 \\ 
			CBD & 0.0050 & 0.0091 & 0.0110 & 0.0005 & 0.0004 & 0 \\ 
			\hline
		\end{tabular}
		\label{tab:RMSE:no_out}
	\end{table}
	
	Table~\ref{tab:QT} reports the averages of the values of the minimized objective functions and the computational times over the 200 simulations. All timings are carried out on a Dell R740 2 x 14 cores (Intel(R) Xeon(R) Gold 6132 CPU @ 2.60GHz) 768 GB RAM 2x200gb SSD 960 GB NVME. CBD performs considerably better in terms of minimizing the objective function than SQN. SQN is somewhat faster than CBD for small $q$ but slower for larger $q$. The computing times for SQN grow quite quickly with increasing $q$ while the computing times seems more stable for CBD.

	The  RMSE results are shown in Table~\ref{tab:RMSE:no_out}. For the calculation of the RMSEs, we exclude small percentages of very extreme parameter estimates. These percentages are reported in the last column of Table~\ref{tab:RMSE:no_out}. CBD performs better than SQN since smaller RMSEs are obtained and there are no outlying parameter estimates. For SQN quite large percentages of extreme parameter estimates are observed.

	\subsection{Assessment of cross-validation and regularization methods with $p=5$} \label{sec:simp5:res}

	In this section we continue with the simulations from the previous setting but restrict attention to CV selection of $q$ and $\lambda$ using CBD for optimization with the LASSO regularization  ($\xi=1$). We select values of $q$ in $\boldsymbol{q}=\{1,2,3,4,5\}$ and values of $\lambda$ in $\boldsymbol{\lambda}=\{0,10^{-3},\ldots,5\}$ which has 20 elements and where the non-zero values of $ \boldsymbol \lambda$ grow log-linearly from $\log 10^{-3}$ to $\log 5$. We consider three situations: (1) we select $q$ from $\boldsymbol{q}$ with $\lambda=0$ fixed, thus least squares estimation (LSE) is performed; (2) we search for the jointly optimal $(q,\lambda)$; (3) we fix $q=5$ and select $\lambda$ from $\boldsymbol{\lambda}$. Recall that the selection of a relatively big $\lambda$ may lead to zero columns in the $\baf$ estimate. We therefore consider the effective $q_{\text{eff}}$ as defined in Section~\ref{sec:cv}. Thereby we can also evaluate the selection of $q$ in situation (3).  In case of (2) we both consider the minimum CV (Min) and the 1-standard error (1-SE) rules to select $q$ and $\lambda$.
	
	Table~\ref{tab:distp5:final} shows the distribution of absolute distance between $q_{\text{eff}}$ and the true $q=2$. For LSE, using the Min rule, $q_{\text{eff}}$ coincides with the true $q$ for 47\% of the simulations and differs at most by 1 from the true $q$ in 75\% of the simulations. The results with the 1-SE rule are similar with percentages $46$ and $78$.
	LASSO with Min rule for  joint selection of $(q,\lambda)$ performs similarly to LSE with the corresponding percentages 42 and 74 \%. With fixed $q=5$ the percentages are reduced to 16\% and 53 \%. Using 1-SE rule, the LASSO forces many columns to be zero leading to quite small percentages where $|q_{\text{eff}}-2| \le 1$.
\begin{table}[!ht]
		\centering
		\renewcommand{\arraystretch}{1.35}
		\setlength{\tabcolsep}{3.5pt}
		\caption{Distribution of $ |q_{\text{eff}}-2|$ (in \%) over 200 simulations from a  multivariate log Gaussian Cox process ($p=5,q=2$) using CBD for minimization. 
		}\label{tab:distp5:final}
		\begin{tabular}{l |cccc | cccc | cccc}
			\hline
			\hline
			\multicolumn{1}{c}{}& \multicolumn{4}{c}{LSE} &  \multicolumn{4}{c}{LASSO} &  \multicolumn{4}{c}{LASSO} \\
			\multicolumn{1}{c}{} & \multicolumn{4}{c}{$q \in \boldsymbol{q}, \;\lambda=0$} &  \multicolumn{4}{c}{$q \in \boldsymbol{q}, \;\lambda \in \boldsymbol{\lambda}$} &  \multicolumn{4}{c}{$q=5\, ;\lambda \in\boldsymbol{\lambda}$} \\
			\hline
			$ |q_{\text{eff}}-2|$ & 0 &1&2&3 & 0 &1&2&3 & 0 &1&2&3 \\ 
			\hline 
			\hline
			Min &47&28&13&12 &42&32&21&5 &16&37&30&17\\ 
			1-SE  &46&32&22&0 &15&20&65&0 &10&22&65&3 \\
			\hline 
		\end{tabular}
	\end{table}
	
	\begin{table}[!ht]
		\centering
		\renewcommand{\arraystretch}{1.35}
		\setlength{\tabcolsep}{4pt}
		\caption{Average  RMSEs obtained from 200 simulations from a multivariate log Gaussian Cox process ($p=5, q=2$) for different methods of selecting $q$ and $\lambda$.}
		\label{tab:RMSEp5:final}
		\begin{tabular}{l rr rr rr rr}
			\hline
			& \multicolumn{2}{c}{$q=2$} & \multicolumn{2}{c}{LSE}  & \multicolumn{2}{c}{LASSO} & \multicolumn{2}{c}{LASSO}  \\
			&  $\lambda=0$ &  $\lambda \in \bld$ & \multicolumn{2}{c}{$q \in \bq, \lambda=0$}  & \multicolumn{2}{c}{$q \in \bq, \lambda \in \bld$} & \multicolumn{2}{c}{$q=5,\lambda \in \bld$}\\
			& Min &  Min  & Min &  1-SE & Min & 1-SE & Min& 1-SE\\
			\hline
			$\hat \baf \hat \baf^\T$ & 0.26 & 0.33 &0.33&0.40 &0.36&0.54 &0.40&0.54\\
			$\hat \bsigma^2$ & 0.42 & 0.54 &0.54&0.58 &0.56&0.75 &0.63&0.76\\
			$\hat \bpsi$& 0.04 & 0.05 &0.05&0.02 &0.03&0.01 &0.04&0.01\\
			$\hat {\mathrm{PV}}$ & 0.28 & 0.31 &0.32& 0.35 &0.33&0.41 &0.37&0.42\\
			\hline
		\end{tabular}
	\end{table}
  
	RMSEs are reported in Table~\ref{tab:RMSEp5:final} for all three situations. In addition, in the first columns, we consider the case fixed $q=2$ assuming the true $q$ is known. We first note that LASSO gives worse results than LSE when $q=2$ is fixed. In general, for unknown $q$,  LSE and LASSO perform quite similarly when the Min rule is used. The results are worse when 1-SE is used and in particular for LASSO. When $q$ is fixed to $5$ and only $\lambda$ is selected the results are worse than for LASSO with $q$ selected by the Min rule while the results with $q=5$ are similar to LASSO with $q$ selected by the 1-SE rule.

 The overall impression is that LSE performs slightly better than LASSO, especially in estimating $\baf \baf^\T$. This may indicate that when $p$ is relatively small, selection of $q$ with $\lambda=0$ (LSE) already gives sparse results. Another reason that LASSO does not improve RMSE may be that the true $\baf$ is not that sparse having only 40\% zero components. Thus the bias introduced by regularization is not counterbalanced by a reduction in variance. Also, the 1-SE rule does not seem preferable in this situation. In the next section we consider a more complex setting with $p=10$.

	\subsection{Assessment of cross-validation and regularization methods with $p=10$} \label{sec:simp10}

	In this experiment, we study a more complex situation with a higher $p$ and more variation in the parameters. We simulate 200 point patterns from a multivariate log Gaussian Cox process  with $p=10$, $q=4$, $W=[0,1]^2$, and parameters
	\begin{align*}
	\boldsymbol{\phi} & = (0.02,0.03,0.03,0.05)^\T, \\
	\boldsymbol{\sigma}^2 & = (1,1,1.5,1,0.2,0.2,1,1.5,1.5,1.5)^\T,
	\end{align*}
	
	\begin{align*}
	\boldsymbol\af & =
	\begin{bmatrix}
	& \sqrt{0.5} & 0.10 & -1& 0 \\ 
	& 0 & 0 & -0.70 & 1 \\ 
	& 0 & -0.15 & \sqrt{0.5} & 0.10 \\ 
	& -1& 0 & 0 & 0 \\ 
	& -0.70 & 1& 0 & -0.15 \\ 
	& \sqrt{0.5} & 0.10 & -1& 0 \\ 
	& 0 & 0 & -0.70 & 1 \\ 
	& 0 & -0.15 & \sqrt{0.5} & 0.10 \\ 
	& -1& 0 & 0 & 0 \\ 
	& -0.70 & 1& 0 & -0.15 \\ 
	\end{bmatrix},
	\end{align*}
	and $\bpsi$ equal to 
	\[ (0.01,0.02,0.02,0.03,0.04,0.04,0.05,0.06,0.06,0.07)^\T. \]
	The settings for the trend models, the kernel estimation and the cross validation are as in the previous simulation study except that $\bq=\{0,\ldots,8\}$. In $\baf$, 40\% of the components are zeros and 20\% are of absolute value less than 0.15. The remaining components have absolute value greater than 0.7.

	Table~\ref{tab:distp10:final} shows the distribution of the absolute distance $|q_{\text{eff}}-4|$ between $q_{\text{eff}}$ and the true $q=4$. Considering first the Min rule, with LSE, $q_{\text{eff}}$ concurs with the true $q$ in 19\% of the simulations and differs at most by 2 from the true $q$ in 58\% of the simulations. The corresponding percentages are 14\%  and 65 \% for LASSO, and 6\% and 41 \% for LASSO with $q=8$ fixed. In this situation, the 1-SE rule seems advantageous for selecting $q$. For example, the percentage of $q_{\text{eff}}$'s which differ from the true $q$ by at most 2 improves from 58\% to 83 \% for LSE, from 65\% to 80 \% for LASSO, and from 41\% to 68 \% for LASSO with fixed $q=8$.
	\begin{table*}[!ht]
		\centering
		\renewcommand{\arraystretch}{1.35}
		\caption{Distribution of $ |q_{\text{eff}}-4|$ from 200 simulations of a  multivariate log Gaussian Cox process ($p=10$ and $q=4$).}
		\label{tab:distp10:final}
		\begin{tabular}{l |rrrrr | rrrrr |rrrrr}
			\hline
			\hline
			\multicolumn{1}{c}{}& \multicolumn{5}{c}{LSE} & \multicolumn{5}{c}{LASSO} & \multicolumn{5}{c}{LASSO}\\
			\multicolumn{1}{c}{} & \multicolumn{5}{c}{$q \in \boldsymbol{q}, \;\lambda=0$} &  \multicolumn{5}{c}{$q \in \boldsymbol{q}, \;\lambda \in \boldsymbol{\lambda}$} &  \multicolumn{5}{c}{$q=8\, ;\lambda=\boldsymbol{\lambda}$} \\
			\hline
			$ |q_{\text{eff}}-4|$ & 0&1&2&3&4 & 0&1&2&3&4& 0&1&2&3&4 \\ 
			\hline
			\hline
			Min &19&21&18&19&23 &14&31&20&19&16 &6&15&20&21&38 \\
			1-SE  &27&36&20&12&5 &22&37&21&8&12 &21&22&25&11&21 \\
			\hline
		\end{tabular}
	\end{table*}

	Table~\ref{tab:RMSEp10:final} details the RMSE results. The superiority of the 1-SE rule when selecting $q$ does not translate into better results in terms of RMSE except for LASSO with fixed $q=8$ where better results are obtained with 1-SE than with Min. The best results are obtained with LASSO using the Min rule for selecting $q$ and $\lambda$. This indicates that regularization is indeed helpful in complex settings with relatively large $p$. 
\begin{table}[!ht]
		\centering
		\renewcommand{\arraystretch}{1.35}
		\caption{Average of RMSEs obtained from 200 simulations from a multivariate log Gaussian Cox process ($p=10, q=4$) for different methods of selecting $q$ and $\lambda$.}
		\label{tab:RMSEp10:final}
		\begin{tabular}{l | cc |  ccH |cc}
			\hline
			&  \multicolumn{2}{c}{LSE} &  \multicolumn{3}{c}{LASSO} &  \multicolumn{2}{c}{$q=8$ (LASSO)} \\ 
			\cline{2-8}
			& Min & 1-SE & Min & 1-SE & SE ($\lambda$ \& $q$) & Min & 1-SE\\
			\hline
			$\hat \baf \hat \baf^\T$  &0.50&0.67&0.44&0.48&&0.78&0.51 \\ 
			$\hat \bsigma^2$  &0.58&0.89&0.54&0.70&&0.88&0.76 \\
			$\hat \bpsi$ &0.02&0.02&0.01&0.02&&0.02&0.02 \\
			$\hat{\mathrm{PV}}$ &0.35&0.35&0.34&0.39&&0.35&0.40\\
			\hline
		\end{tabular}
	\end{table}

	Based on the simulation studies,  for analyzing highly multivariate point pattern data, we recommend to use regularization with the Min rule for selecting $q$ and $\lambda$.

	\section{Application} \label{sec:app}
	In a 50-hectare $1,000 \; \mathrm{m} \times  500 \; \mathrm{m}$ region of the tropical moist forest of Barro Colorado Island (BCI) in central Panama, censuses have been carried out where all free-standing woody stems with at least 10 mm diameter at breast height were identified, tagged, and mapped, resulting in maps of over 350,000 individual trees with around 300 species \citep[see e.g.][]{hubbell:foster:83,condit:hubbell:foster:96,condit:98}. In addition, 13 spatial covariates are also available containing topological attributes and soil nutrients (see Figure~\ref{fig:cov}). Our main objective is to study the impact of regularization and the computational feasibility of our method.
	We first consider  9 tree species, {\em Psychotria, Protium t., Capparis, Protium p., Swartzia, Hirtella, Tetragastris, Garcinia, Mourmiri}, with intermediate abundances ranging from 2500 to 7500 and previously analyzed by \cite{waagepetersen:etal:16}. The plots of locations of each species are shown in Figure~\ref{fig:spec9}. The main aim of this analysis is to compare the results with our new algorithm to those obtained by \cite{waagepetersen:etal:16}. Secondly, to test our algorithm in a more challenging situation, we analyze a highly multivariate point pattern involving species of trees with at least 400 individuals, resulting in 86 species. 

For each species, we use maximum composite likelihood to fit log-linear regression models involving the spatial covariates for the $\mu_i$-terms in \eqref{eq:Z}. We then estimate the cross pair correlation function using \eqref{eq:ghat}. Therefore, the variation due to observed covariates are filtered out and the non-parametric estimates of cross pair correlation function hence capture the residual correlation due to unobserved covariates, species-specific factors, and any other sources. 

	\subsection{Application with 9 species}
\begin{figure*}[!ht]
	 	\renewcommand{\arraystretch}{0}
	 	\setlength{\tabcolsep}{0pt}
	 	\begin{tabular}{lll}
	 		 \includegraphics[width=0.33\textwidth]{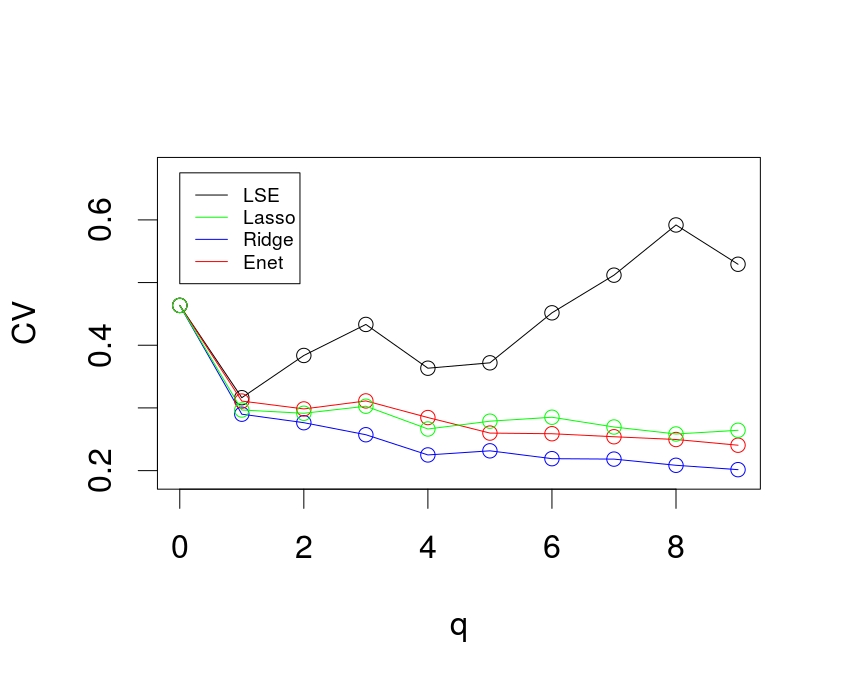} & \includegraphics[width=0.33\textwidth]{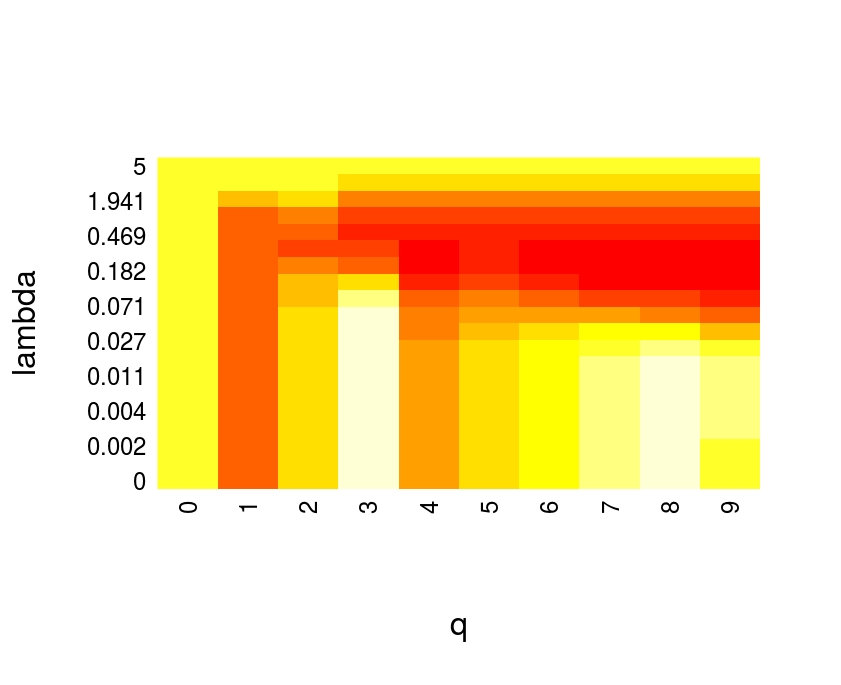} & 
	 		\includegraphics[width=0.33\textwidth]{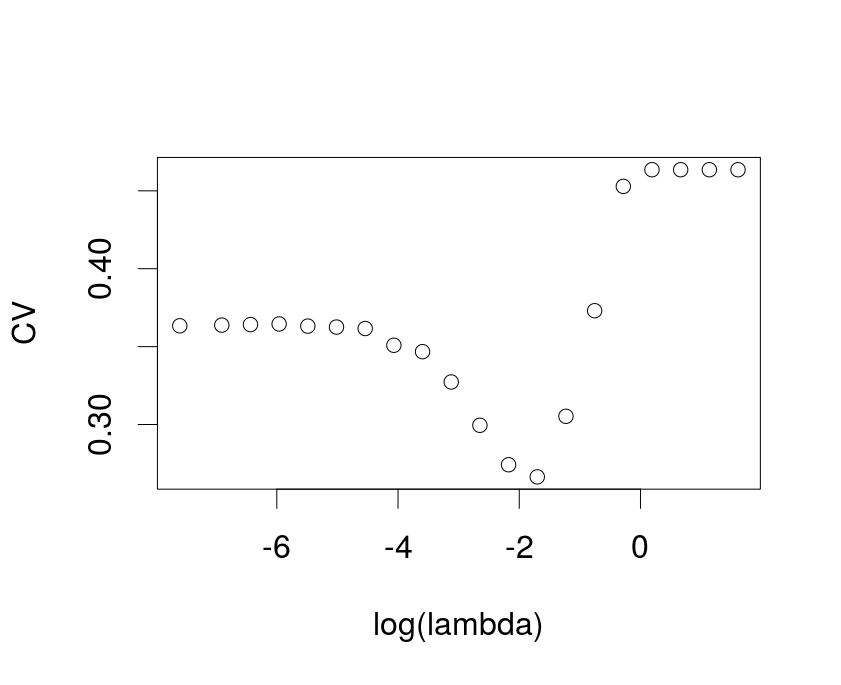} \\
	 	\end{tabular}
	 	\caption{Cross-validation (CV) scores for 9-species data analysis. Left:  $\min_{\ld \in \bld} \mathrm{CV}(q,\ld)$ against $q$ for ridge, elastic net and LASSO and  $\mathrm{CV}(q,0)$ against $q$ for LSE. Middle: image plot of $\mathrm{CV}(q,\lambda)$ in case of ridge (darker color corresponds to smaller CV score). Right: $\mathrm{CV}(4,\lambda)$ plotted against $\log \lambda$.}
	 	\label{fig:CV9}
	 \end{figure*}

	For each value of $\xi=0,0.5,1$ we apply $8$-fold CV to select  $q$ and $\ld$ where $\ld \in \bld= \{0,10^{-3},\ldots,5\}$  as in the simulation studies and $q \in \bq=\{0,\ldots,9\}$. The upper left plot in Figure~\ref{fig:CV9} shows for each $\xi$, $\min_{\ld \in \bld} \mathrm{CV}(q,\ld)$  as a function of $q$. For comparison with \cite{waagepetersen:etal:16} we also show in this plot $\mathrm{CV}(q,0)$ against $q$ (LSE).  
	A general pattern for ridge, elastic net and LASSO is that the cross validation scores decrease quite quickly as a function of $q$ until around $q=4$ and after that the CV scores stabilize or decrease slowly. The CV scores for ridge ($\xi=0$) are consistently smaller than those for elastic net ($\xi=0.5$) and LASSO ($\xi=1$). Hence we select $\xi=0$. The minimal CV score for $\xi=0$ is obtained with $q=9$. However, in the interest of model simplicity, we choose $q=4$ and $\lambda=0.29$ since the decrease in CV score is rather minor from $q=4$ to $q=9$.

	 For comparison, the minimal CV score with LASSO is obtained with $q=8$ and $\ld=0.11$. However, in this case, the resulting effectively selected $q_{\text{eff}}$ is three since the resulting estimate of $\baf$  has 5 zero columns. In case of LSE ($\lambda=0$), the CV procedure chooses $q=1$. The second-smallest CV with LSE is obtained with $q=4$ which was the value chosen in \cite{waagepetersen:etal:16}. The difference in cross validation results for LSE compared with \cite{waagepetersen:etal:16} is due to our new more efficient optimization algorithm, cf.\ the comparison in Section~\ref{sec:simp5:study}.
	 
	 The middle plot in Figure~\ref{fig:CV9} is an image plot of the CV scores for ridge ($\xi=0$) where darker color corresponds to smaller CV score. The development of the CV scores across values of $q$ for fixed $\ld$ appears quite erratic with several local minima. In contrast, for each $q$ there appears to be a well-defined minimum for $\lambda$. As an example, the right plot in Figure~\ref{fig:CV9} shows $\mathrm{CV}(4,\lambda)$ plotted against $\log \lambda$ (where we replace the undefined $\log 0$ by $\log 5e-4$).  The computing time required to run the CV method with $\xi=0$ is $2.4$ hours with the same processor as used in the simulation study.  Approximately $16$ seconds is required to estimate the parameters for the 9-species application using ridge with $q=4$ and $\ld=0.29$.
	 
	 \begin{figure*}[!ht]
	 	\renewcommand{\arraystretch}{0}
	 	\setlength{\tabcolsep}{0pt}
	 	\begin{tabular}{ll}
	 		\includegraphics[width=0.5\textwidth]{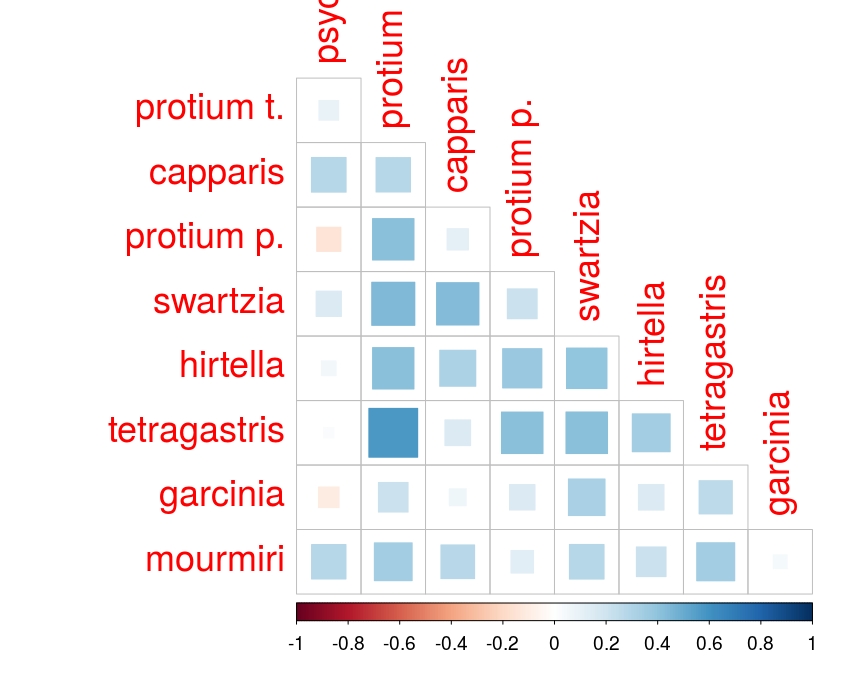}	
	 		& \includegraphics[width=0.5\textwidth]{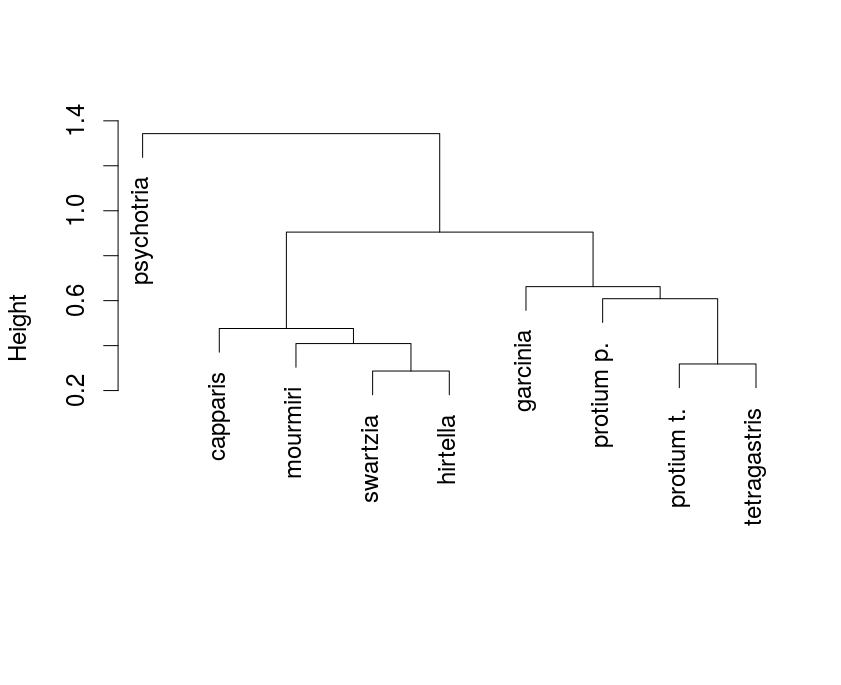} \\
	 	\end{tabular}
	 	\caption{Left: Estimated inter-species correlations ${\mathrm{corr}}\{Z_i(\bu), Z_j(\bu)\}$ at lag zero. Right: 9-species clustering based on $\|\hat \baf_{i.}- \hat \baf_{j.} \|$.}
	 	
	 	\label{fig:mult9}
	 \end{figure*}

	The results regarding the multivariate dependence structure of the 9 species are qualitatively similar to those obtained by \cite{waagepetersen:etal:16}. 
	The estimated inter-species correlations $\mathrm{corr}\{Z_i(u),Z_j(u)\}$, cf.\ \eqref{eq:covZ}, are shown in the left plot of  Figure~\ref{fig:mult9}. Most of the pairs of species have a positive correlation. However, the correlations between {\em Psychotria} and the other species are mainly close to zero. 
The right plot in Figure~\ref{fig:mult9} shows a hierarchical clustering of the species based on the estimated coefficient rows $\baf_{i \cdot}$, where {\em Psychotria} appears to form its own cluster in agreement with the estimated inter-species correlations. 
This clustering may have some relation to the families of species as shown by the cluster of {\em Protium p.}, {\em Protium t.} and {\em Tetragastris} which come from the same family (see Table~\ref{supp-tab:86spec} in the supplementary material). 

	\subsection{Application with 86 tree species}
	
\begin{figure*}[h]
	\renewcommand{\arraystretch}{0}
	\setlength{\tabcolsep}{0pt}
	\begin{tabular}{lll}
		\includegraphics[width=0.333\textwidth]{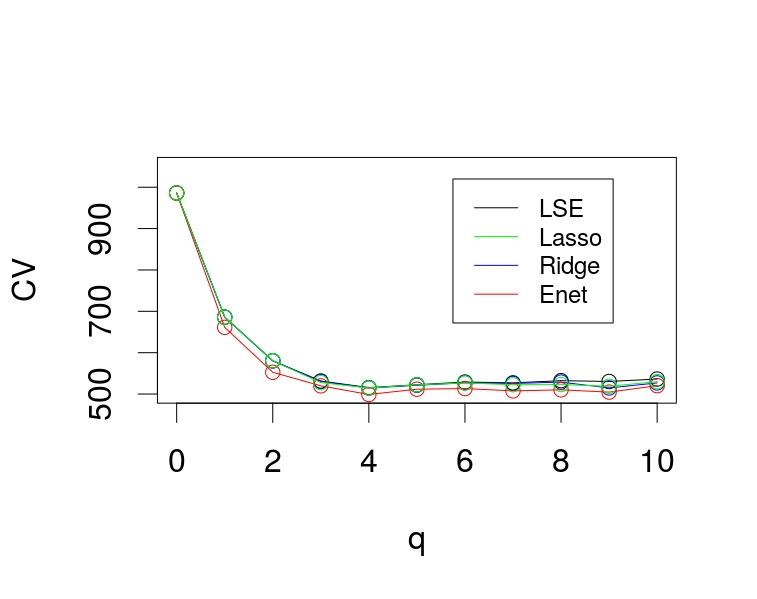} & \includegraphics[width=0.333\textwidth]{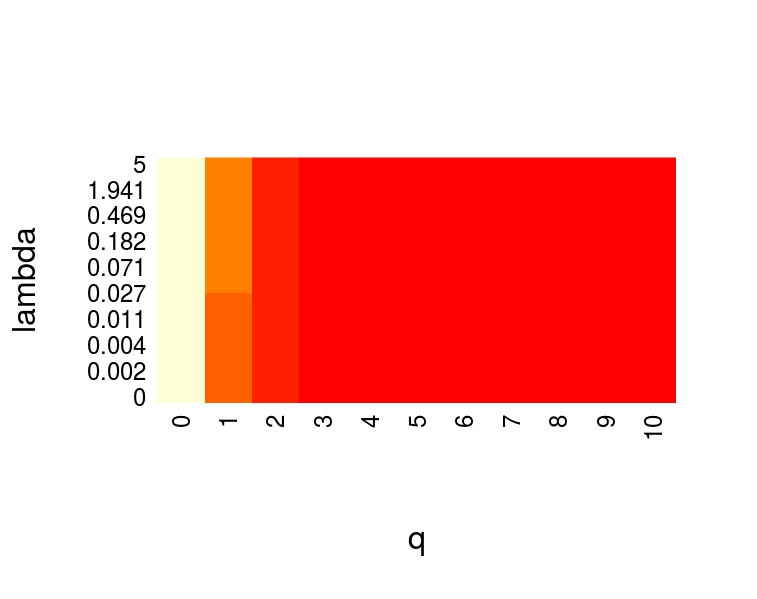} &
		\includegraphics[width=0.333\textwidth]{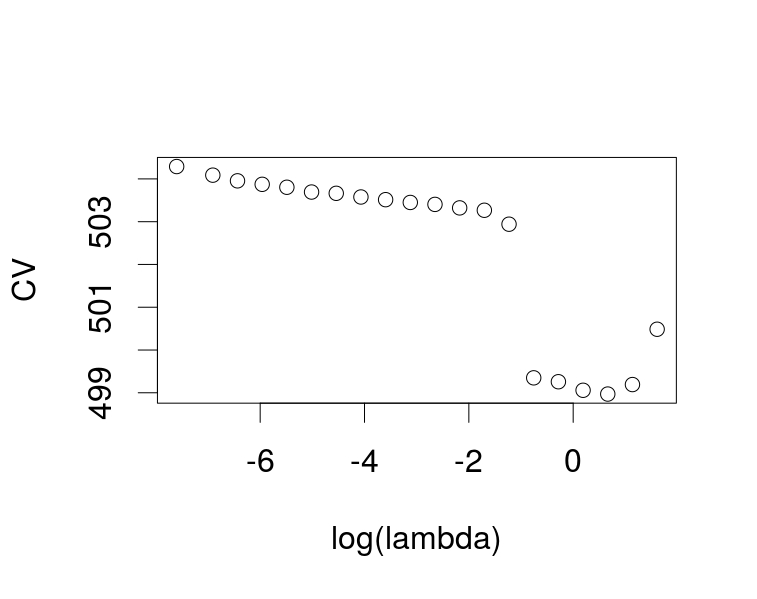}
	\end{tabular}
	\caption{CV scores for 86-species data analysis. Left:  $\min_{\ld \in \bld} \mathrm{CV}(q,\ld)$ against $q$ for ridge, elastic net and LASSO and  $\mathrm{CV}(q,0)$ against $q$ for LSE. Middle: image plot of $\mathrm{CV}(q,\lambda)$ in case of elastic net (darker color corresponds to smaller CV score). Right: $\mathrm{CV}(4,\lambda)$ plotted against $\log \lambda$.}
	\label{fig:CV86}
\end{figure*}

For the 86-species application, we apply the 8-fold CV procedure with $\xi=0,0.5,1$ and $\ld \in \{0,10^{-3},\ldots,5\}$ as in the previous section and $q \in \{0, \ldots, 10\}$. Figure~\ref{fig:CV86} is similar to Figure~\ref{fig:CV9}. The left plot shows that consistently smaller CV scores are obtained with elastic net ($\xi=0.5$) and the smallest CV score is obtained with $q=4$. The remaining plots are obtained with $\xi=0.5$. The image plot of cross validation scores in the middle plot looks much smoother than in the 9 species case. The right plot shows a well defined minimum for $\lambda=1.94$ given $q=4$.

\begin{table}[h]
	\centering
	\caption{Distribution (in \%) of estimated inter-species correlations $\mathrm{corr}[Y_i(u),Y_j(u)]$ and $\mathrm{corr}[Z_i(u),Z_j(u)]$, $i \neq j$, over different intervals $[\text{Lower},\text{Upper}]$ for the 86 species application using elastic net ($\xi=0.5$) with $q=4$ and $\ld=1.94$. 
	}\label{tab:correlation}
	\renewcommand{\arraystretch}{1.35}
	\begin{tabular}{rrrrrrr}
		\hline
		Lower	& -1   & -0.5    & -0.2  & 0    & 0.2   & 0.5 \\ 
		Upper &  -0.5 &  -0.2  &  0 &  0.2  &  0.5  &   1\\ 
		\hline
		$\mathrm{corr}[Y_i(u),Y_j(u)]$ & 2 & 6 & 9 & 13 & 22 & 48 \\ 
		$\mathrm{corr}[Z_i(u),Z_j(u)]$  & 0 & 2 & 15 & 60 & 19 & 4 \\ 
		\hline
	\end{tabular}
\end{table}

\begin{table}[!ht]
	\centering
	\caption{Distribution of estimated $\mathrm{PV}_i(0)$ for 86 species application using elastic net ($\xi=0.5$) with $q=4$ and $\ld=1.94$.}\label{tab:PV}
	\renewcommand{\arraystretch}{1.35}
	\begin{tabular}{rrrrr}
		\hline
		Interval & 0-0.25 & 0.25-0.5 & 0.5-0.75 & 0.75-1 \\ 
		\hline
		Number of species & 46 & 20 & 10 & 10 \\ 
		Species (\%) & 53 & 23 & 12 & 12 \\ 
		\hline
	\end{tabular}
\end{table}

The computing time for the CV is 7.6 hours for $\xi=0.5$ and the computing time to estimate the parameters for the chosen $q=4$ and $\lambda=1.94$ is 3.2 minutes. Out of $4\times 86$ parameters in the estimated $\baf$, 13 were set to zero by the elastic net regularization. We thereby model $86\times 87/2=3741$ distinct pair and cross pair correlation functions using only $6\times 86-13+4=507$ parameters. Thus we have indeed obtained a sparse model for the given data.

The distribution of estimated PVs is shown in Table~\ref{tab:PV}. Most species ($53\%$) have estimated proportions of variances due to common factors less than 0.25.

Table~\ref{tab:correlation} shows the distribution of estimated inter-species correlations due to common latent fields and the combination of common and species-specific fields (see Section~\ref{sec:multvarinference}) across 6 intervals. Most estimated correlations are positive. However, the correlations decrease a lot in absolute value when the species-specific fields are included (last row of Table~\ref{tab:correlation}).

\begin{figure*}[!ht]
	\renewcommand{\arraystretch}{0}
	\setlength{\tabcolsep}{0pt}
	\centering
	\includegraphics[width=1\textwidth, height=0.8\textwidth]{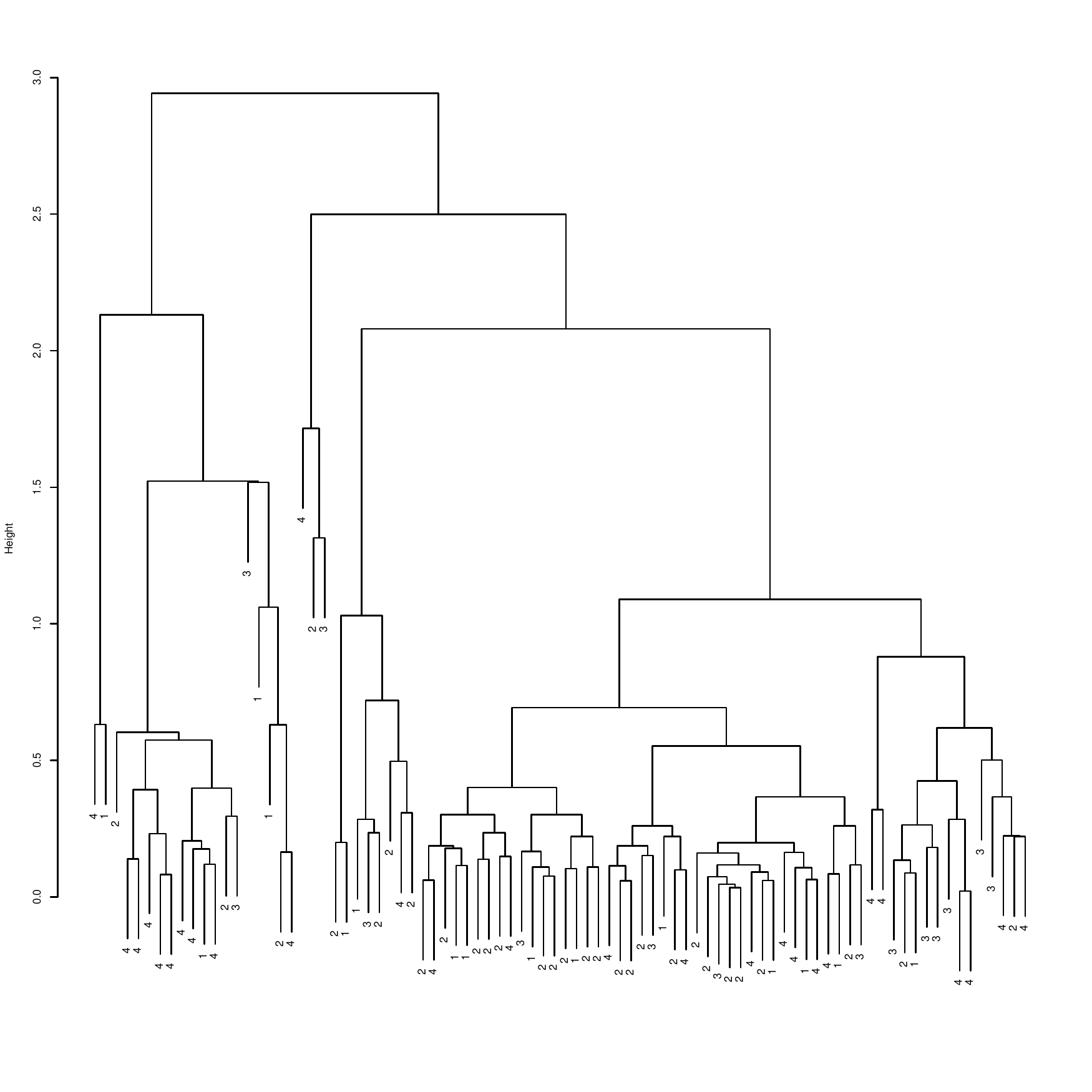} 
	\caption{86-species clustering based on $\|\hat \baf_{i.}- \hat \baf_{j.} \|$. Leaves are marked with species life form [1] Shrub, [2] Understory, [3] Midstory, and [4] Tree.}
	\label{fig:clus86}
\end{figure*}

Figure~\ref{fig:clus86} shows a clustering of species based on estimated $\baf_{i\cdot}$, $i=1,\ldots,86$. The leaves are marked with species life form. There may be some indication that species of life form ``Tree'' (life form number 4) tend to cluster together. However, one should be careful with this interpretation since apparent patterns like this could be due to sampling variation.

	\section{Conclusion} \label{sec:disc}
	
	We developed in this study a regularized estimation method for highly multivariate point patterns modeled by multivariate log Gaussian Cox processes. The procedure is numerically stable and performs well both in the considered simulations and applications.  In our truly highly multivariate second application, we were able to fit a sparse model for a multivariate point pattern with 86 types of points.

	An interesting application of obtained estimates is to group types of points according to their estimated dependence on common latent fields as expressed by the rows $\baf_{i\cdot}$. Hence a further development could be to consider an extension of the so-called fused LASSO \citep{Tibshirani05sparsityand} by introducing regularization for differences $\baf_{i \cdot}-\baf_{j\cdot}$. A further possibility would be to consider a sparse group LASSO \citep{simon2013sparse} to obtain estimates of $\baf$ with some zeros of $\alpha_{il}$ as developed in this paper and, in addition, with entire rows of zeros implying independence of corresponding types of points and all other types of points.

	\vspace{1cm}
	{\bf Acknowledgements} The research by A.\ Choiruddin,  F.\ Cuevas-Pacheco, and R.\ Waagepetersen is supported by The Danish Council for Independent Research | Natural Sciences, grant DFF – 7014-00074 "Statistics for point processes in space and beyond", and by the "Centre for Stochastic Geometry and Advanced Bioimaging", funded by grant 8721 from the Villum Foundation.
		
		The BCI forest dynamics research project was made possible by National
	Science Foundation grants to Stephen P. Hubbell: DEB-0640386,
	DEB-0425651, DEB-0346488, DEB-0129874, DEB-00753102, DEB-9909347,
	DEB-9615226, DEB-9615226, DEB-9405933, DEB-9221033, DEB-9100058,
	DEB-8906869, DEB-8605042, DEB-8206-992, DEB-7922197, support from the
	Center for Tropical Forest Science, the Smithsonian Tropical Research k+1
	Institute, the John D. and Catherine T. MacArthur Foundation, the
	Mellon Foundation, the Celera Foundation, and numerous private
	individuals, and through the hard work of over 100 people from 10
	countries over the past two decades. The plot project is part of the Center for Tropical Forest Science, a global network of large-scale demographic tree plots.
	
	The BCI soils data set were collected and analyzed by J.\ Dalling,
	R.\ John, K.\ Harms, R.\ Stallard and J.\ Yavitt  with support from
	NSF DEB021104, 021115, 0212284, 0212818 and OISE 0314581, STRI and
	CTFS. Paolo Segre and Juan Di Trani provided assistance in the
	field. The covariates \texttt{dem}, \texttt{grad}, \texttt{mrvbf},
	\texttt{solar} and \texttt{twi} were computed in SAGA GIS by Tomislav
	Hengl (\texttt{http://spatial-analyst.net/}). 
	We thank Dr.\ Joseph
	Wright for sharing data on dispersal modes and life forms for the BCI
	tree species.
	
	\bibliographystyle{plainnat}
	\bibliography{masterbib}

\begin{thebibliography}{26}
\providecommand{\natexlab}[1]{#1}
\providecommand{\url}[1]{\texttt{#1}}
\expandafter\ifx\csname urlstyle\endcsname\relax
  \providecommand{\doi}[1]{doi: #1}\else
  \providecommand{\doi}{doi: \begingroup \urlstyle{rm}\Url}\fi

\bibitem[Baddeley et~al.(2014)Baddeley, Jammalamadaka, and
  Nair]{baddeley:jammalamadaka:nair:14}
Adrian Baddeley, Aruna Jammalamadaka, and Gopalan Nair.
\newblock Multitype point process analysis of spines on the dendrite network of
  a neuron.
\newblock \emph{Journal of the Royal Statistical Society: Series C (Applied
  Statistics)}, 63\penalty0 (5):\penalty0 673--694, 2014.

\bibitem[Chil{\` e}s and Delfiner(1999)]{chiles:delfiner:99}
Jean-Paul Chil{\` e}s and Pierre Delfiner.
\newblock \emph{Geostatistics: modeling spatial uncertainty}.
\newblock Probability and Statistics. Wiley, New York, 1999.

\bibitem[Choi et~al.(2010)Choi, Oehlert, and Zou]{choi2010penalized}
Jang Choi, Gary Oehlert, and Hui Zou.
\newblock A penalized maximum likelihood approach to sparse factor analysis.
\newblock \emph{Statistics and its Interface}, 3\penalty0 (4):\penalty0
  429--436, 2010.

\bibitem[Choiruddin et~al.(2018)Choiruddin, Coeurjolly, and
  Letu{\'e}]{choiruddin2018convex}
Achmad Choiruddin, Jean-Fran{\c{c}}ois Coeurjolly, and Fr{\'e}d{\'e}rique
  Letu{\'e}.
\newblock Convex and non-convex regularization methods for spatial point
  processes intensity estimation.
\newblock \emph{Electronic Journal of Statistics}, 12\penalty0 (1):\penalty0
  1210--1255, 2018.

\bibitem[Coeurjolly et~al.(2017)Coeurjolly, M{\o}ller, and
  Waagepetersen]{coeurjolly:moeller:waagepetersen:17}
Jean-Fran{\c{c}}ois Coeurjolly, Jesper M{\o}ller, and Rasmus Waagepetersen.
\newblock A tutorial on {P}alm distributions for spatial point processes.
\newblock \emph{International Statistical Review}, 85\penalty0 (3):\penalty0
  404--420, 2017.

\bibitem[Condit(1998)]{condit:98}
R.~Condit.
\newblock \emph{Tropical Forest Census Plots}.
\newblock Springer-Verlag and R.\ G.\ Landes Company, Berlin, Germany and
  Georgetown, Texas, 1998.

\bibitem[Condit et~al.(1996)Condit, Hubbell, and
  Foster]{condit:hubbell:foster:96}
Richard Condit, Stephen~P Hubbell, and Robin~B Foster.
\newblock Changes in tree species abundance in a neotropical forest: impact of
  climate change.
\newblock \emph{Journal of tropical ecology}, 12\penalty0 (2):\penalty0
  231--256, 1996.

\bibitem[Diggle et~al.(2005)Diggle, Zheng, and Durr]{diggle:etal:05}
Peter Diggle, Pingping Zheng, and Peter Durr.
\newblock Nonparametric estimation of spatial segregation in a multivariate
  point process: bovine tuberculosis in {C}ornwall, {UK}.
\newblock \emph{Journal of the Royal Statistical Society: Series C (Applied
  Statistics)}, 54\penalty0 (3):\penalty0 645--658, 2005.
\newblock ISSN 1467-9876.
\newblock \doi{10.1111/j.1467-9876.2005.05373.x}.
\newblock URL \url{http://dx.doi.org/10.1111/j.1467-9876.2005.05373.x}.

\bibitem[Friedman et~al.(2010)Friedman, Hastie, and
  Tibshirani]{friedman:hastie:tibshirani:10}
Jerome Friedman, Trevor Hastie, and Robert Tibshirani.
\newblock Regularization paths for generalized linear models via coordinate
  descent.
\newblock \emph{Journal of Statistical Software}, 33\penalty0 (1):\penalty0
  1--22, 2010.

\bibitem[Hastie et~al.(2015)Hastie, Tibshirani, and
  Wainwright]{hastie:tibshirani:wainwright:15}
T.~Hastie, R.~Tibshirani, and M.~Wainwright.
\newblock \emph{Statistical learning with sparsity: the lasso and
  generalizations}.
\newblock Monographs on Statistics and Applied Probability. Chapman \& Hall/CRC
  Press, Boca Raton, 2015.

\bibitem[Hastie et~al.(2013)Hastie, Tibshirani, and
  Friedman]{hastie:tibshirani:friedman:13}
Trevor Hastie, Robert Tibshirani, and Jerome Friedman.
\newblock \emph{The elements of statistical learning: data mining, inference,
  and prediction}.
\newblock Springer Series in Statistics. Springer New York Inc., New York, 2
  edition, 2013.

\bibitem[Hoerl and Kennard(1988)]{hoerl1988ridge}
Arthur~E Hoerl and Robert~W Kennard.
\newblock Ridge regression.
\newblock \emph{Encyclopedia of {S}tatistical {S}ciences}, 8, 1988.

\bibitem[Hubbell and Foster(1983)]{hubbell:foster:83}
S.~P. Hubbell and R.~B. Foster.
\newblock Diversity of canopy trees in a neotropical forest and implications
  for conservation.
\newblock In S.~L. Sutton, T.~C. Whitmore, and A.~C. Chadwick, editors,
  \emph{Tropical Rain Forest: Ecology and Management}, pages 25--41. Blackwell
  Scientific Publications, Oxford, 1983.

\bibitem[Jalilian et~al.(2015)Jalilian, Guan, Mateu, and
  Waagepetersen]{jalilian:etal:15}
A.~Jalilian, Y.~Guan, J.~Mateu, and R.~Waagepetersen.
\newblock Multivariate product-shot-noise {C}ox models.
\newblock \emph{Biometrics}, 71\penalty0 (4):\penalty0 1022--1033, 2015.

\bibitem[Lee et~al.(2014)Lee, Sun, and Saunders]{lee:sun:saunders:14}
Jason~D. Lee, Yuekai Sun, and Michael~A. Saunders.
\newblock Proximal {N}ewton-type methods for minimizing composite functions.
\newblock \emph{SIAM Journal on Optimization}, 24\penalty0 (3):\penalty0
  1420--1443, 2014.

\bibitem[M{\o}ller and Waagepetersen(2003)]{moeller:waagepetersen:03}
J.~M{\o}ller and R.~Waagepetersen.
\newblock \emph{Statistical inference and simulation for spatial point
  processes}.
\newblock Chapman and Hall/CRC, Boca Raton, 2003.

\bibitem[M{\o}ller et~al.(1998)M{\o}ller, Syversveen, and
  Waagepetersen]{moeller:syversveen:waagepetersen:98}
J.~M{\o}ller, A.~R. Syversveen, and R.~Waagepetersen.
\newblock Log {G}aussian {C}ox processes.
\newblock \emph{Scandinavian Journal of Statistics}, 25\penalty0 (3):\penalty0
  451--482, 1998.

\bibitem[M{\o}ller and Waagepetersen(2007)]{moeller:waagepetersen:07}
Jesper M{\o}ller and Rasmus Waagepetersen.
\newblock Modern statistics for spatial point processes.
\newblock \emph{Scandinavian Journal of Statistics}, 34\penalty0 (4):\penalty0
  643--684, 2007.

\bibitem[Rajala et~al.(2018)Rajala, Murrell, and
  Olhede]{rajala:murrell:olhede:17}
Tuomas Rajala, D.~J. Murrell, and S.~C. Olhede.
\newblock Detecting multivariate interactions in spatial point patterns with
  {G}ibbs models and variable selection.
\newblock \emph{Journal of the Royal Statistical Society: Series C (Applied
  Statistics)}, 67\penalty0 (5):\penalty0 1237--1273, 2018.

\bibitem[Simon et~al.(2013)Simon, Friedman, Hastie, and
  Tibshirani]{simon2013sparse}
Noah Simon, Jerome Friedman, Trevor Hastie, and Robert Tibshirani.
\newblock A sparse-group lasso.
\newblock \emph{Journal of Computational and Graphical Statistics}, 22\penalty0
  (2):\penalty0 231--245, 2013.

\bibitem[Thurman et~al.(2015)Thurman, Fu, Guan, and
  Zhu]{thurman2015regularized}
Andrew~L Thurman, Rao Fu, Yongtao Guan, and Jun Zhu.
\newblock Regularized estimating equations for model selection of clustered
  spatial point processes.
\newblock \emph{Statistica Sinica}, 25\penalty0 (1):\penalty0 173--188, 2015.

\bibitem[Tibshirani(1996)]{tibshirani:96}
R~Tibshirani.
\newblock Regression shrinkage and selection via the lasso.
\newblock \emph{Journal of the Royal Statistical Society: Series B (Statistical
  Methodology)}, 58\penalty0 (1):\penalty0 267--288, 1996.

\bibitem[Tibshirani et~al.(2005)Tibshirani, Saunders, Rosset, Zhu, and
  Knight]{Tibshirani05sparsityand}
Robert Tibshirani, Michael Saunders, Saharon Rosset, Ji~Zhu, and Keith Knight.
\newblock Sparsity and smoothness via the fused lasso.
\newblock \emph{Journal of the Royal Statistical Society: Series B (Statistical
  Methodology)}, 67\penalty0 (1):\penalty0 91--108, 2005.

\bibitem[Waagepetersen(2007)]{waagepetersen:07}
R.~Waagepetersen.
\newblock An estimating function approach to inference for inhomogeneous
  {N}eyman-{S}cott processes.
\newblock \emph{Biometrics}, 63\penalty0 (1):\penalty0 252--258, 2007.

\bibitem[Waagepetersen et~al.(2016)Waagepetersen, Guan, Jalilian, and
  Mateu]{waagepetersen:etal:16}
R.~Waagepetersen, Y.~Guan, A.~Jalilian, and J.~Mateu.
\newblock Analysis of multi-species point patterns using multivariate log
  {G}aussian {C}ox processes.
\newblock \emph{Journal of the Royal Statistical Society: Series C (Applied
  Statistics)}, 65\penalty0 (1):\penalty0 77--96, 2016.

\bibitem[Zou and Hastie(2005)]{zou:hastie:05}
H.~Zou and T.~Hastie.
\newblock Regularization and variable selection via the elastic net.
\newblock \emph{Journal of the Royal Statistical Society: Series B (Statistical
  Methodology)}, 67\penalty0 (2):\penalty0 301--320, 2005.

\end{thebibliography}
	
	\newcommand{\paartial}{\dd}
	
	\appendix

	\section{Proximal Newton Method} \label{prox}
	Suppose we want to find the solution of
	\begin{align}
	\min_{\boldsymbol{\theta} \in \mathbb{R}^n} f(\boldsymbol{\theta}):= a(\boldsymbol{\theta}) + c(\boldsymbol{\theta}),
	\end{align} 
	where the function $f(\cdot)$ can be separated into two parts: the function $a(\cdot)$ which is a convex and twice continuously differentiable loss function and the function $c(\cdot)$ which is a convex but not necessarily differentiable penalty function. The proximal-Newton method is an iterative optimization algorithm that uses a quadratic approximation of the differentiable part $a(\cdot)$: 
	\begin{align} 
	f(\boldsymbol{\theta})  \approx & \; \hat f(\boldsymbol{\theta}) \nonumber \\
	 = & \; \hat a(\boldsymbol{\theta}) + c(\boldsymbol{\theta}) \nonumber \\
	 = & \; a(\boldsymbol{\ta^{(k)}}) + \nabla a(\boldsymbol{\ta^{(k)}})^\T (\boldsymbol{\theta} - \boldsymbol{\ta^{(k)}}) + (\boldsymbol{\theta} - \boldsymbol{\ta^{(k)}})^\T H(\boldsymbol{\ta^{(k)}}) (\boldsymbol{\theta} - \boldsymbol{\ta^{(k)}}) +  c(\boldsymbol{\theta}) \label{proxmethod},
	\end{align}
	where $\bta^{(k)}$ is the current value of $\bta$, $\nabla a(\cdot)$ is the first derivative of $a(\cdot)$ and $H(\cdot)$ is an approximation to the Hessian matrix $\nabla^2 a(\cdot)$. Letting $\tilde \bta = \argmin_{\bta} \hat f(\bta)$, the next value of $\bta$ is obtained as
	\[ \bta^{(k+1)}=\bta^{(k)}+ t (\tilde \bta -  \bta^{(k)}) \]
	for some $t>0$. That is, $\tilde \bta$ is used to construct a search direction for the $k+1$th value of $\bta$. Theoretical results in \cite{lee:sun:saunders:14} show that $t$ can be chosen so that $f(\bta^{(k+1)})<f(\bta^{(k)})$. The matrix $H(\cdot)$ can be chosen in various ways, see \cite{lee:sun:saunders:14} and \cite{hastie:tibshirani:wainwright:15} for more details.
	
	In the following sections, we adapt the proximal Newton method to minimization of our objective function.

	\subsection{Quadratic approximation for updating $\baf_{i\cdot}$}\label{taylor}
	Let us first regard \eqref{eq:blockdescent} as a function of $\baf_{i \cdot}$, 
	\begin{align}
	Q_{\ld,i}(\baf_{i \cdot},\sigma^2_i)  = & \; 2 \sum_{\substack{j=1\\j \neq i}}^p \| Y_{ij}-
	\tilde X_{ij}\baf_{i \cdot} \|^2 +  \|Y_{ii}- X_{ii} \bbt_{ii}(\baf,\bsigma^2) \|^2 + \ld \sum_{l=1}^q  p(\af_{il}) \nonumber \\
	= & \;  a(\baf_{i \cdot}) + b(\baf_{i \cdot}) + c(\baf_{i \cdot}).\label{eq:abc}
	\end{align}
	
	To minimize \eqref{eq:blockdescent}, we consider the proximal Newton method stated in \eqref{proxmethod}. In particular, we approximate $b(\baf_{i \cdot})$ by a quadratic approximation around the current value $\baf_{i \cdot}^{(k)}$:
	\begin{align}
	b(\baf_{i \cdot}) \approx & \; \hat b(\baf_{i \cdot}) \nonumber\\
	= & \; b(\baf_{i \cdot}^{(k)})+ \nabla b(\baf_{i \cdot}^{(k)})^\T  (\baf_{i\cdot} -\baf_{i \cdot}^{(k)}) + \frac{1}{2} (\baf_{i\cdot}-\baf_{i \cdot}^{(k)})^\T H(\baf_{i \cdot}^{(k)}) (\baf_{i\cdot} -\baf_{i \cdot}^{(k)}) \label{approxb}.
	\end{align}
Here, the first derivative is
	\[
	\nabla b(\baf_{i \cdot}^{(k)})= -4 D(\baf_{i \cdot}^{(k)}) X_{ii,\cdot(1:q)}^\T \Big(Y_{ii}- X_{ii} \bbt_{ii}(\baf^{(k)},\bsigma^2)\Big) 
\]
while $H(\baf_{i \cdot}^{(k)})$ is an approximation of the second derivative,
\[
	\nabla^2 b(\baf_{i \cdot}^{(k)})= 8 D(\baf_{i \cdot}^{(k)}) X_{ii,\cdot(1:q)}^\T X_{ii,\cdot(1:q)} D(\baf^{(k)}_{i \cdot}) - C(\baf^{(k)}_{i \cdot}),
\]
	where $D(\baf_{i \cdot}^{(k)})=\diag(\af_{i1}^{(k)},\ldots,\af_{iq}^{(k)})$, $X_{ii,\cdot (1:q)}$ denotes the first $q$ columns in $X_{ii}$, and $C(\baf^{(k)}_{i \cdot})=4 \diag\bigg(X_{ii,\cdot (1:q)}^\T \Big(Y_{ii}- X_{ii} \bbt_{ii}(\baf^{(k)},\bsigma^2)\Big)\bigg)$. Specifically,
	\begin{align*}
		H(\baf_{i \cdot}^{(k)}) & = 8 D(\af^{(k)}_{i \cdot}) X_{ii,\cdot(1:q)}^\T X_{ii,\cdot(1:q)} D(\af^{(k)}_{i \cdot}) \\
		& \approx \nabla^2 b(\baf_{i \cdot}^{(k)}).
	\end{align*}
	
	To ease the presentation and computation, we write $\hat b(\baf_{i \cdot})$ from \eqref{approxb} in the form of a least squares problem
	\begin{align*}
	\hat b(\baf_{i \cdot}) = & \; \|Y_{ii}- X_{ii} \bbt_{ii}(\baf^{(k)},\bsigma^2) \|^2 \\ &- 2  \Big(Y_{ii}- X_{ii} \bbt_{ii}(\baf^{(k)},\bsigma^2) )\Big)^\T [2X_{ii,\cdot (1:q)} D(\baf_{i \cdot}^{(k)})] (\baf_{i\cdot} -\baf_{i \cdot}^{(k)}) \\
	& + \frac{1}{2} (2) (\baf_{i\cdot} -\baf_{i \cdot}^{(k)})^\T [2 D(\baf_{i \cdot}^{(k)}) X_{ii,\cdot (1:q)}^\T] [2X_{ii,\cdot (1:q)} D(\baf_{i \cdot}^{(k)})] (\baf_{i\cdot} -\baf_{i \cdot}^{(k)}) \\
	= & \; \mathbf{v}^\T \mathbf{v} - 2\mathbf{v}^\T X^*_{ii}\boldsymbol{\gamma} + \boldsymbol{\gamma}^\T (X^*_{ii})^\T X^*_{ii}\boldsymbol{\gamma}\\
	= & \; \|\mathbf{v} - X^*_{ii}\boldsymbol{\gamma}\|^2 \\
	= & \; \|Y^*_{ii} - X^*_{ii} \baf_{i \cdot}\|^2
	\end{align*}
	where
	\begin{align*}
	\mathbf{v} & = Y_{ii}- X_{ii} \bbt_{ii}(\baf^{(k)},\bsigma^2) ,\\
	X^*_{ii} & = 2X_{ii,\cdot (1:q)} D(\baf_{i \cdot}^{(k)}), \\
	\boldsymbol{\gamma} & = \baf_{i\cdot} -\baf_{i \cdot}^{(k)}, \\
	Y^*_{ii} & = Y_{ii}+ X_{ii,\cdot (1:q)}\baf_{i \cdot}^{2,(k)} - X_{ii,\cdot (q+1)}\sigma^{2}_i.
	\end{align*}
	Replacing $b$ in \eqref{eq:abc} with $\hat b$ we obtain the approximate objective function $\hat Q_{\ld,i}(\baf_{i \cdot}|\baf_{i \cdot}^{(k)})$ given in \eqref{finalobj}. Since \eqref{finalobj} is a standard regularized  least squares problem, we minimize \eqref{finalobj} using a coordinate descent algorithm to obtain $\hat \baf_{i \cdot}$ as detailed in Section~\ref{alpha}. 
	
	\subsection{Theoretical result regarding proximal Newton update}\label{theory}

	Let $\Delta(\baf_{i \cdot}^{(k)})=\hat \baf_{i \cdot}-\baf_{i \cdot}^{(k)}$ where $\hat \baf_{i \cdot}$ is the minimizer of \eqref{finalobj} and according to a line search strategy let
	\[   \baf_{i \cdot}^{(k+1)}= \baf_{i \cdot}^{(k)}+ t\Delta(\baf_{i \cdot}^{(k)}) \]
	for some $t>0$.
	Following the proof of Proposition~2.3 in \cite{lee:sun:saunders:14}, we can verify the following theorem.
	\begin{theorem}\label{descentdirection}
		Let $ H(\baf_{i \cdot}^{(k)})=8 D(\baf_{i \cdot}^{(k)}) X_{ii}^\T X_{ii} D(\baf_{i \cdot}^{(k)})$. Then
		\begin{align*}
		Q_{i,\ld}(\baf_{i \cdot}^{(k+1)},\sigma^2_i) \leq & \; Q_{i,\ld} (\baf_{i \cdot}^{(k)},\sigma^2_i) - t \Delta (\baf_{i \cdot}^{(k)})^\T H(\baf_{i \cdot}^{(k)}) \Delta (\baf_{i \cdot}^{(k)}) + O(t^2).
		\end{align*}
	\end{theorem}
	Thus, by Theorem~\ref{descentdirection}, if $H(\baf_{i \cdot}^{(k)})$ is positive definite, we can choose $t>0$ so that  $Q_{i,\ld}(\baf_{i \cdot}^{(k+1)},\sigma^2_i)< Q_{i,\ld}(\baf_{i \cdot}^{(k)},\sigma^2_i)$. That is, the update of $\baf_{i \cdot}$ results in a decrease of the objective function \eqref{eq:blockdescent}.
	
	\section{Algorithm} \label{sec:CDA}
	
	In our block descent algorithm, we minimize \eqref{eq:reglso} with
	respect to $ \bsigma^2, \baf, \bphi$, and $\bpsi$ in turn. For $i=1,\ldots,p$, we first update $\sigma^2_i$ by minimizing \eqref{eq:blockdescent} using least squares estimation followed by an update of
	$\baf_{i\cdot}$ by minimizing \eqref{finalobj} using a coordinate descent method.  We denote by $X_{ij,\cdot k}$ the $k$th column of $X_{ij}$ for $k=1,\ldots,q$ ($i \neq j$) or $k=1,\ldots,q+1$ ($i=j$). We detail, respectively in Appendices~\ref{sigma} and \ref{alpha}, the updates of $\sigma^2_i$ and the coordinate descent updates of $\af_{il}$ for $l=1,\ldots,q$. A summary of the final algorithm is given by Appendix~\ref{algorithm}.
	
	\subsection{Update of $\sigma_i^2$} \label{sigma}
	
	The parameter $\hat \sigma_i^2$ is updated using least squares methods. More precisely, the gradient of \eqref{eq:blockdescent} with respect to $\sigma_i^2$ is
	\begin{align*}
	\frac{\partial Q_{\ld,i}(\baf_{i \cdot},\sigma_i^2)}{\partial \sigma_i^2} = & \; -2 
	X_{ii,\cdot (q+1)}^\T (Y_{ii}- X_{ii} \bbt_{ii}(\baf,\bsigma^2) ).
	\end{align*}
	By solving $\frac{\partial Q_{\ld,i}(\baf_{i \cdot},\sigma_i^2)}{\partial \sigma_i^2} = 0$, we obtain the update
	\begin{align}
	\label{updatesigma}
	\sigma_i^2 \leftarrow \max \left\{    \frac{
		X_{ii,\cdot (q+1)}^\T \left (Y_{ii}- \sum_{l=1}^q X_{ii,\cdot l} \af^2_{i l}
		\right )
	}{X_{ii,\cdot (q+1)}^{\T} X_{ii,\cdot (q+1)}}  ,0 \right \} 
	\end{align}
	where $\max\{a,0\}$ is used to avoid negative results of the update.
	
	\subsection{Update of $\af_{il}$} \label{alpha}
	Let $  r_{ij}= Y^*_{ij} - \sum_{\substack{k=1\\k \neq l}}^q X^*_{ij,\cdot k} \af_{ik}$, where $Y^*_{ij}$ and $X^*_{ij}$ are specified in \eqref{eq:YstarXstar}.
	Then we rewrite \eqref{finalobj} as
	\begin{align*}
	\hat Q_{\ld,i}(\baf_{i\cdot}) = & \; \sum_{\substack{j=1}}^p \|r_{ij}- X^*_{ij,\cdot l} \af_{il}\|^2 + \ld \sum_{\substack{k=1\\k \neq l}}^q \Big( (1-\xi)\frac{1}{2}\af_{ik}^2 + \xi | \af_{ik}|\Big) \\ & + \ld \Big((1-\xi)\frac{1}{2}\af_{il}^2 + \xi | \af_{il}| \Big).
	\end{align*}
	The gradient with respect to $\af_{il}$ is
	\begin{align*}
	\frac{\partial \hat Q_{\ld,i}(\af_{il})}{\partial \af_{il}} & =  \; -2 \sum_{\substack{j=1}}^p (X^*_{ij,\cdot l})^\T (r_{ij}- X^*_{ij,\cdot l} \af_{il}) + \ld \Big((1-\xi)\af_{il} + \xi \sign(\af_{il}) \Big).
	\end{align*}
	Following the main argument by \cite{friedman:hastie:tibshirani:10}, the coordinate-wise update for $\af_{il}$ is of the form
	\begin{align}
	\label{updatealpha}
	\af_{il} \leftarrow
	\frac{S \left (2 \sum_{\substack{j=1}}^p (X^*_{ij,\cdot l})^\T r_{ij},\ld \xi \right)}{2 \sum_{\substack{j=1}}^p (X^*_{ij,\cdot l})^\T X^*_{ij,\cdot l} + \ld(1-\xi)},
	\end{align}
	where $ S(A,\ld \xi) =\text{sign}(A)(|A|-\ld \xi)_+$.
		
	\subsection{Algorithm to update $\baf, \bsigma^2, \bphi, \bpsi$} \label{algorithm}
	For a given $q$ and sequence of $\lambda$ values $0 \le \ld_1,\ldots,\ld_M$, the overall procedure to estimate the parameters: $\baf, \bsigma^2, \bphi, \bpsi$ is described  by Algorithm~\ref{Algorithm:BDA}. Note that estimates obtained with $\ld_{s-1}$ are used as initial values for the estimation with $\ld_{s}$, $s=2,\ldots,M$.
	
	\alglanguage{pseudocode}
	\begin{algorithm}[!ht]
		\small
		\caption{Cyclical block descent method for minimization of regularized least squares objective function \eqref{eq:reglso}.}
		\label{Algorithm:BDA}
		\begin{algorithmic}[!ht]
			\State 
			\State Set initial values $\hat \baf^{(0)}, \hat \bsigma^{2,(0)},\hat \bphi^{(0)}$ and $\hat \bpsi^{(0)}$
			\For{$s=1 \mbox{ to } M$}
			\State $\bsigma^2 :=\hat \bsigma^{2,(s-1)}$
			\State $\baf:=\hat \baf^{(s-1)}$
			\State $\bphi:=\hat \bphi^{(s-1)}$
			\State $\bpsi:=\hat \bpsi^{(s-1)}$
			\While{Relative function convergence not achieved}
			
			\For{$i=1 \mbox{ to } p$}
			\State Update  $\sigma_i^2$ using \eqref{updatesigma}
			\State Update $\baf_{i\cdot}$ using cyclical descent over  $\af_{il}$, $l=1,\ldots,q$ using \eqref{updatealpha}
                        \State Apply line search for $\baf_{i\cdot}$ 
			\EndFor
			\State update $\bphi$ using quasi-Newton
			\State update $\bpsi$ using quasi-Newton
			\EndWhile
			\State $\hat \bsigma^{2,(s)}:=\bsigma^2$
			\State $\hat \baf^{(s)}:=\baf$
			\State $\hat \bphi^{(s)}:=\bphi$
			\State $\hat \bpsi^{(s)}:=\bpsi^2$
			\EndFor
		\end{algorithmic}
	\end{algorithm}
	
	\section{Plots and detail information of BCI data used in the analysis} \label{app:bci}
\begin{figure*}[!ht]
		\renewcommand{\arraystretch}{0}
		\setlength{\tabcolsep}{0pt}
		\begin{tabular}{lll}
			\includegraphics[width=0.33\textwidth]{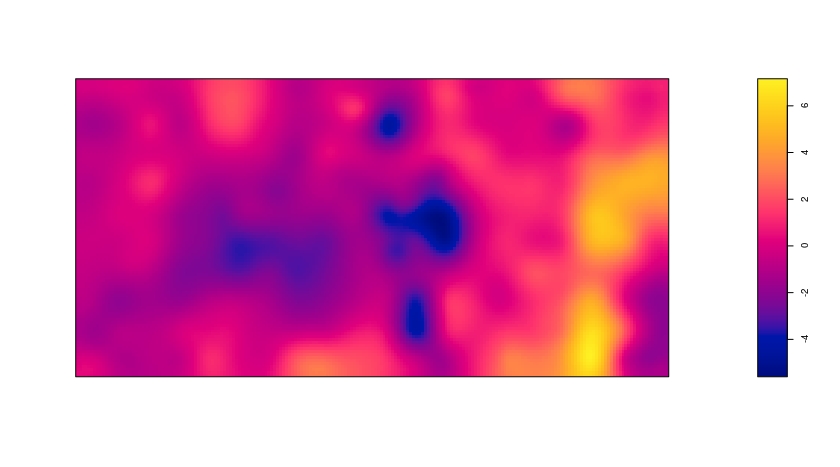} &	\includegraphics[width=0.33\textwidth]{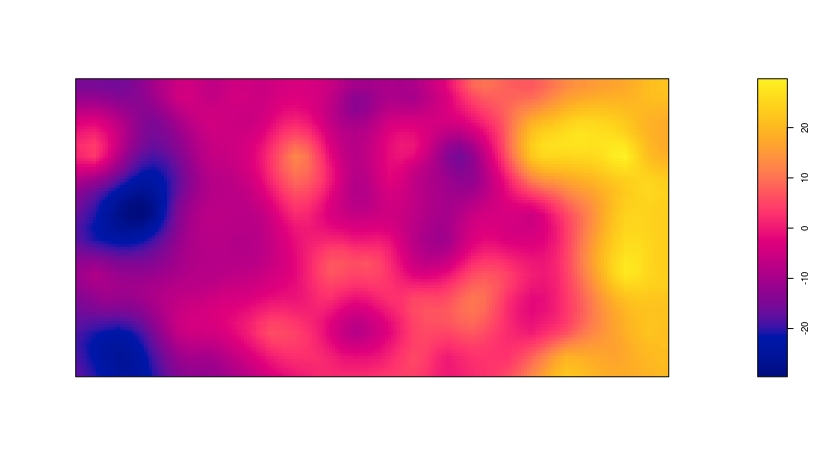} &
			\includegraphics[width=0.33\textwidth]{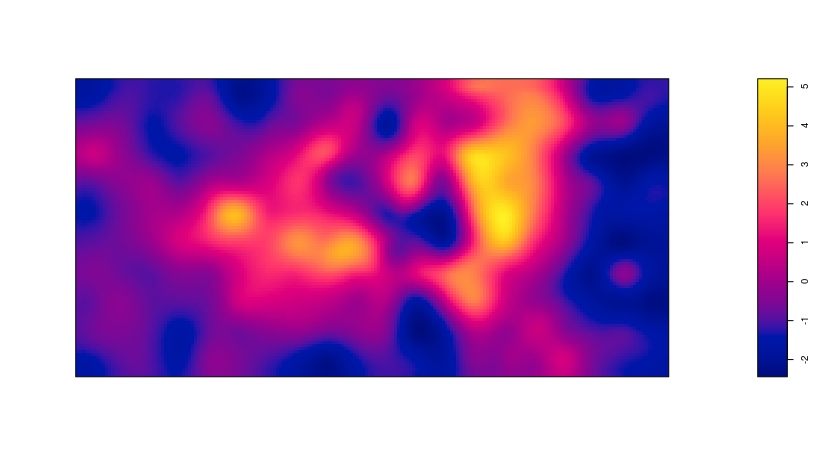} \\
			\includegraphics[width=0.33\textwidth]{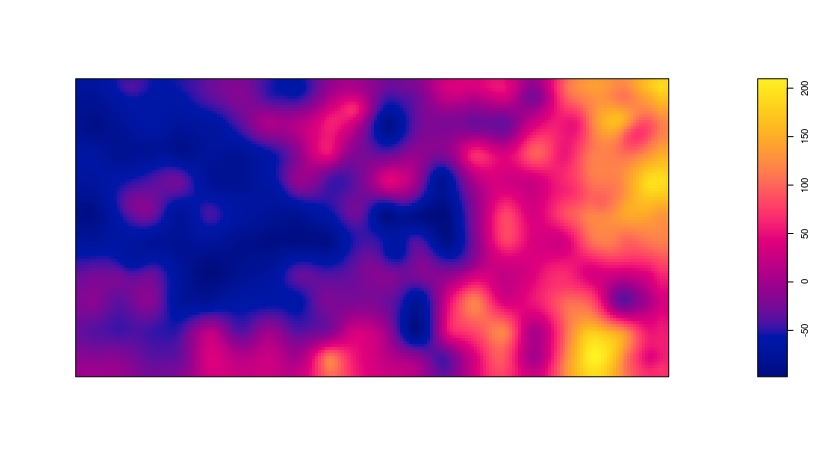} & \includegraphics[width=0.33\textwidth]{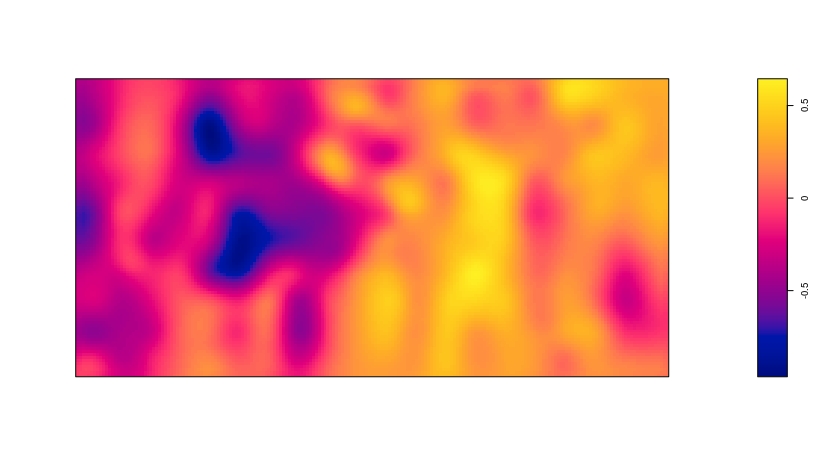} &
			\includegraphics[width=0.33\textwidth]{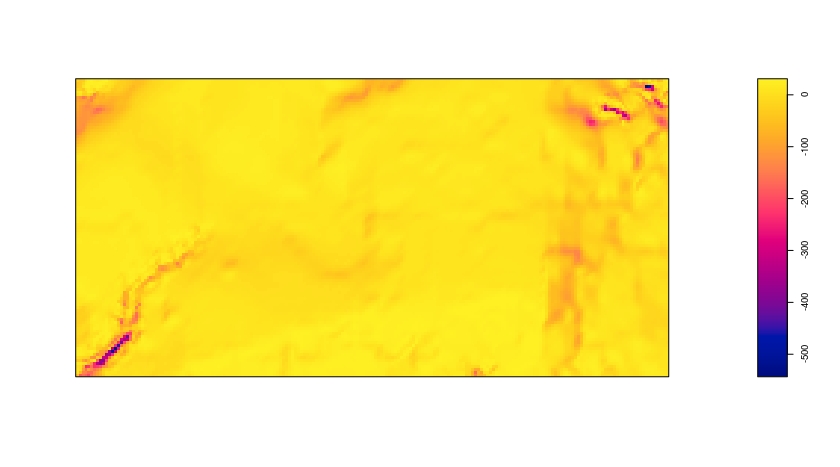} \\
			\includegraphics[width=0.33\textwidth]{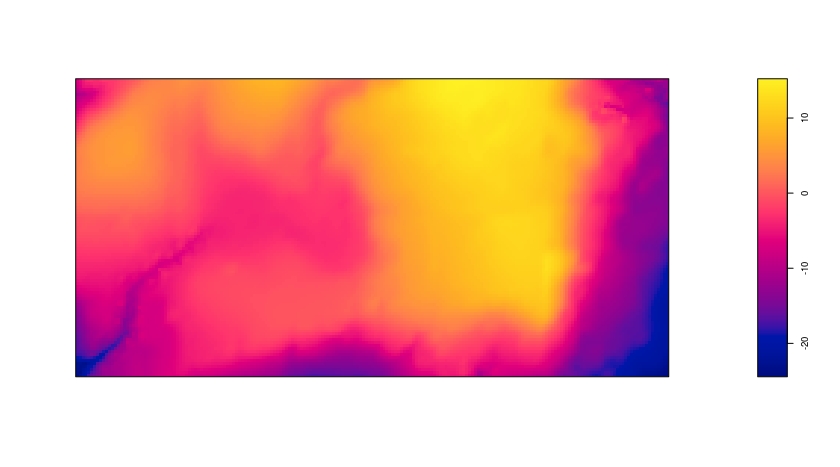} &	\includegraphics[width=0.33\textwidth]{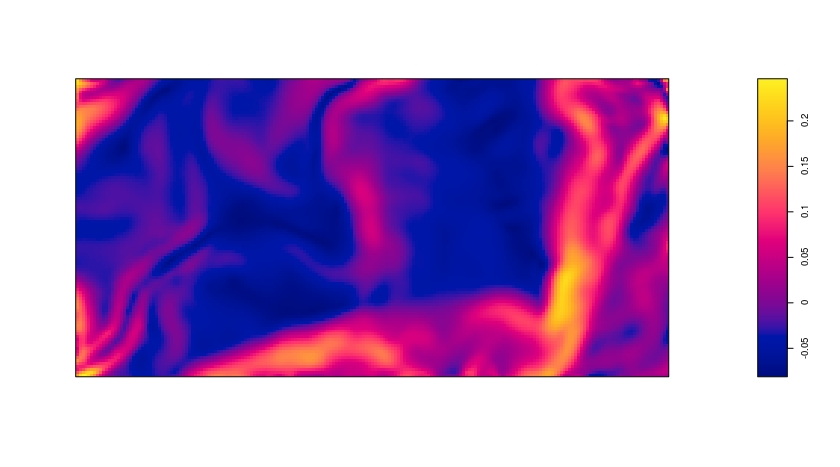} &
			\includegraphics[width=0.33\textwidth]{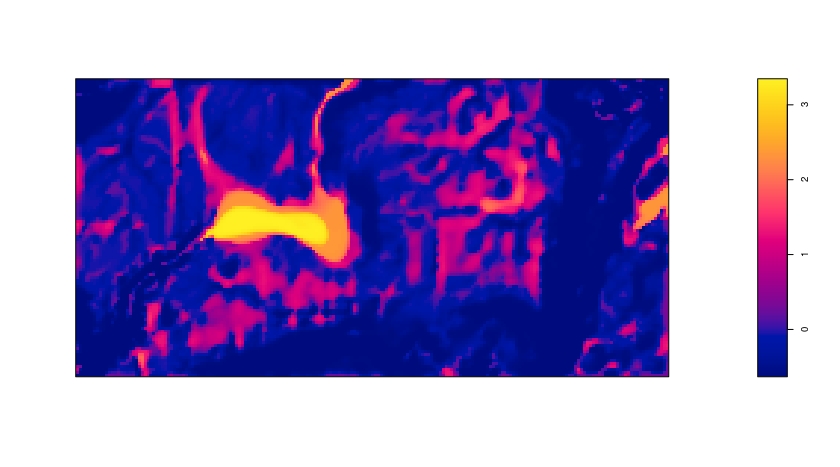} \\
			\includegraphics[width=0.33\textwidth]{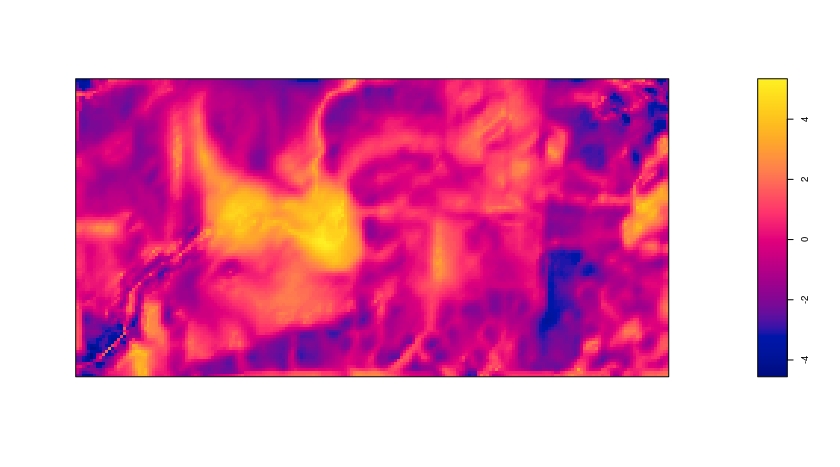} &	\includegraphics[width=0.33\textwidth]{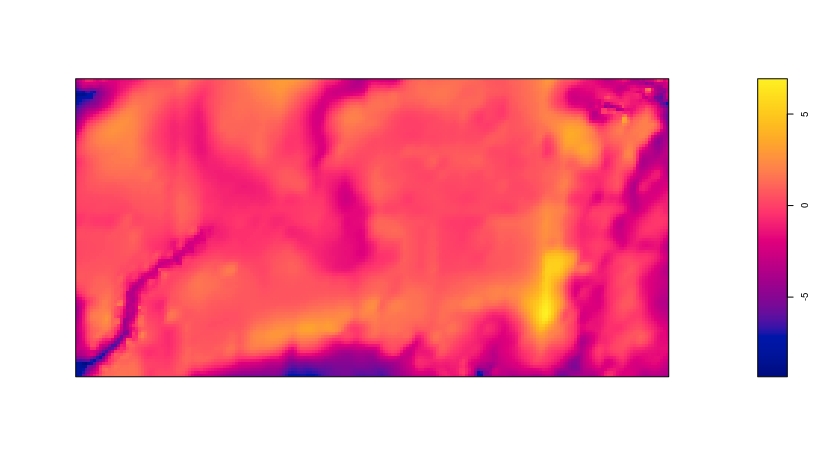} &
			\includegraphics[width=0.33\textwidth]{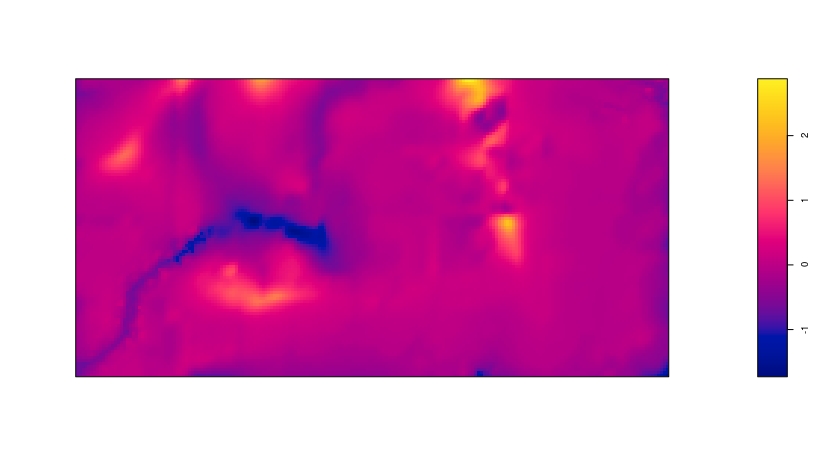} \\
			\includegraphics[width=0.33\textwidth]{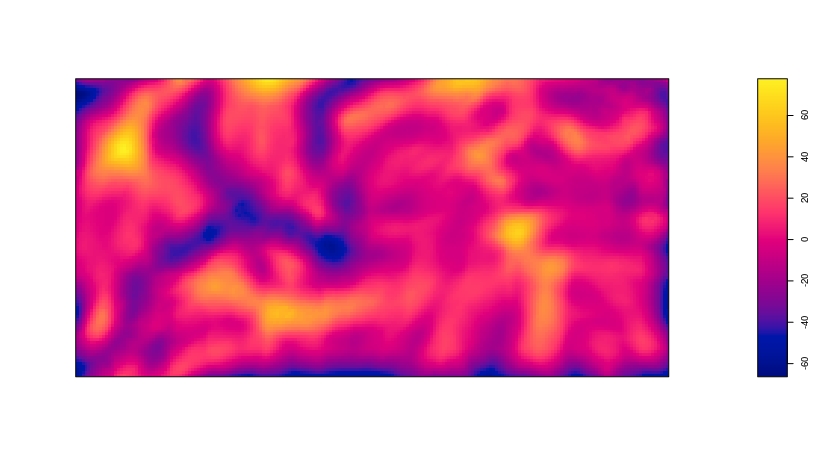} && \\
		\end{tabular}
		\caption{Covariates involved in the analysis (from left to right): 1st row:  Copper content (mg/kg of soil) in the surface soil, mineralization needs for Nitrogen (mg/kg of soil) after a 30-day incubation period and  Phosphorus content (mg/kg of soil) in the surface soil; 2nd row:  Potassium content (mg/kg of soil) in the surface soil,  pH content in the surface soil, and incoming mean annual solar radiation; 3rd row: elevation, slope, and multiresolution index of valley bottom flatness; 4th row: topographic wetness index, difference from the mean value in 15 pixels search radius, and deviation from mean value in 15 pixels search radius; 5th row: convergence index (search radius) with direction to the center cell.}
		\label{fig:cov}
	\end{figure*}

\begin{figure*}[!ht]
		\renewcommand{\arraystretch}{0}
		\setlength{\tabcolsep}{0pt}
		\begin{tabular}{lll}
			\includegraphics[width=0.33\textwidth]{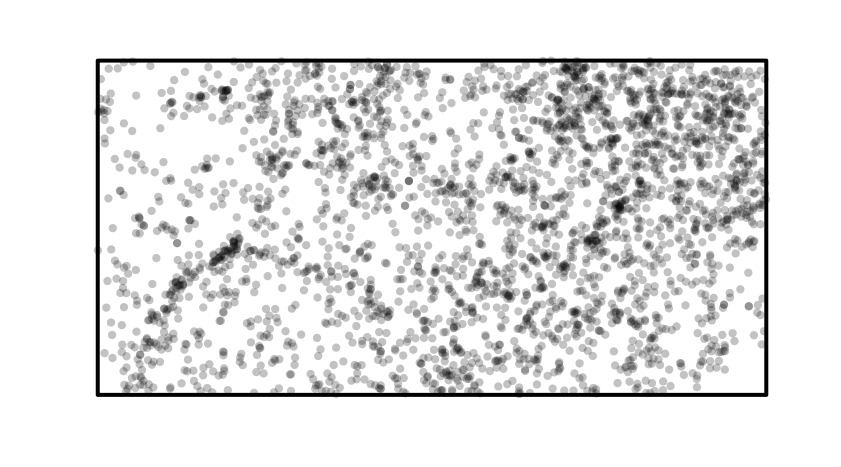} &
			\includegraphics[width=0.33\textwidth]{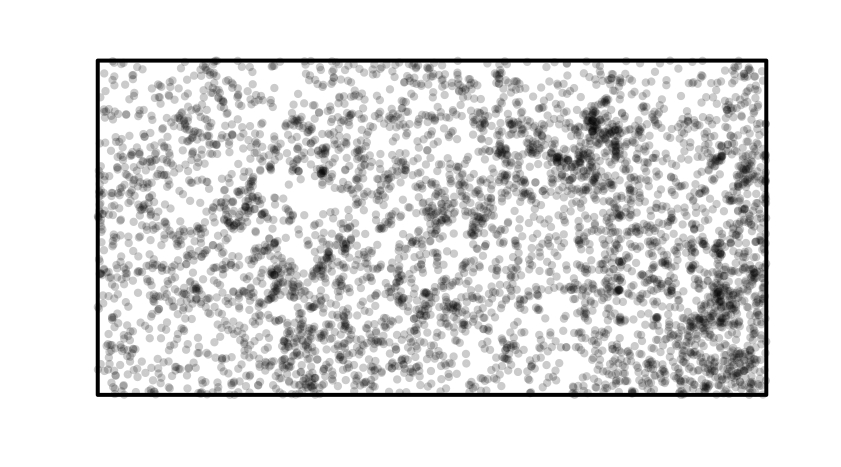} &	\includegraphics[width=0.33\textwidth]{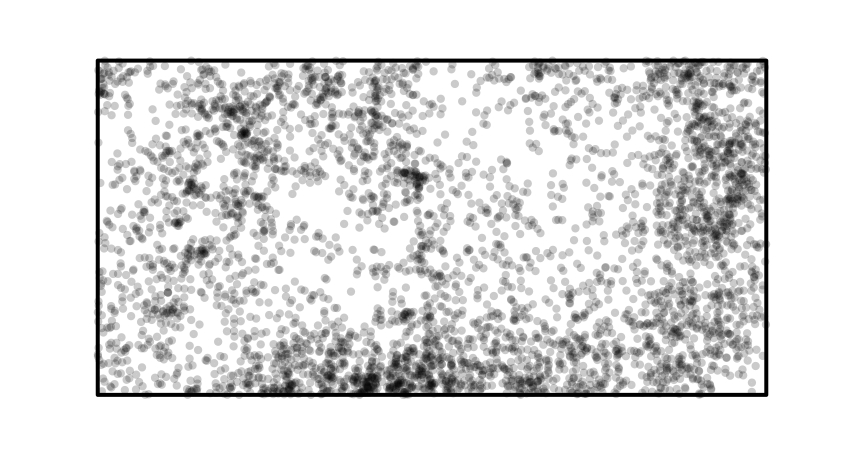} \\
			\includegraphics[width=0.33\textwidth]{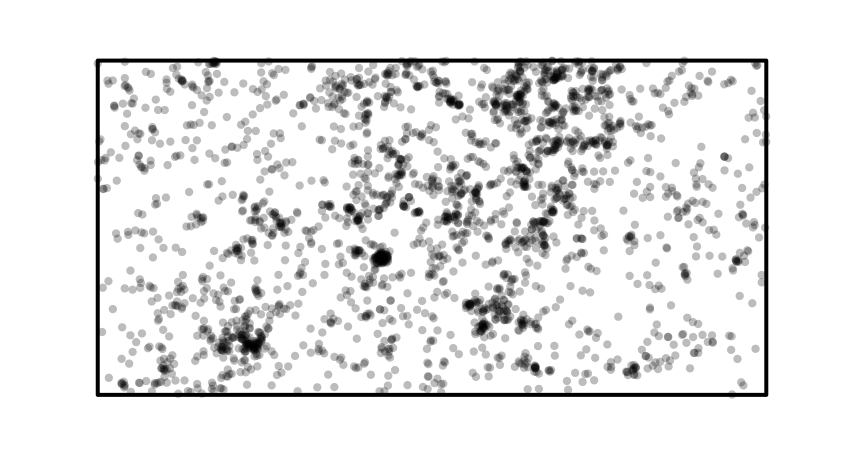} &	\includegraphics[width=0.33\textwidth]{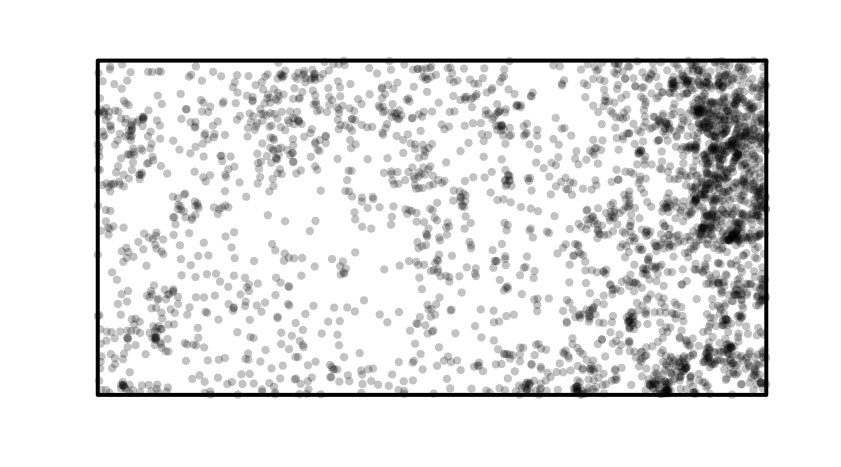} &
			\includegraphics[width=0.33\textwidth]{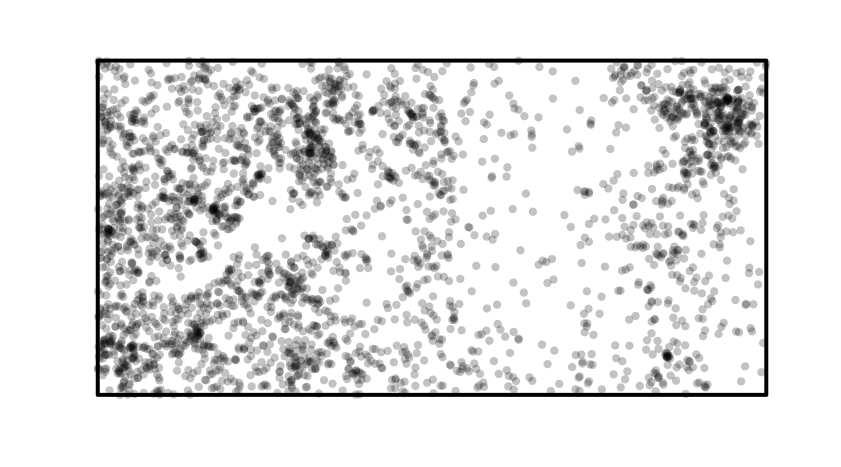} \\
			\includegraphics[width=0.33\textwidth]{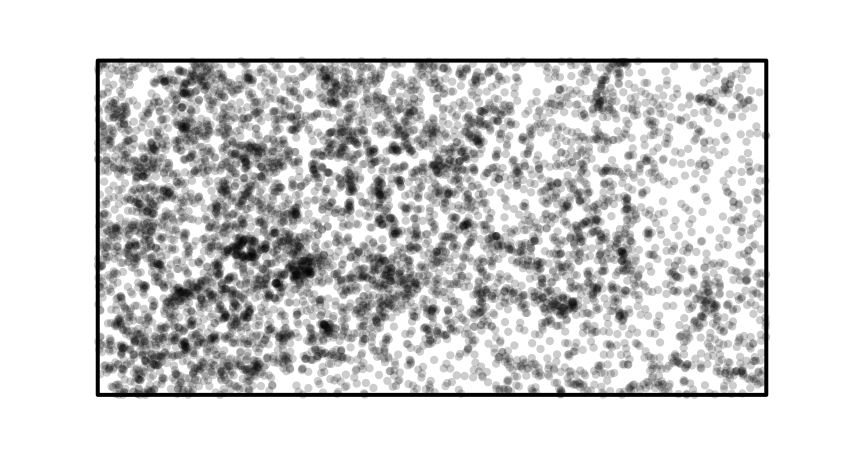} &
			\includegraphics[width=0.33\textwidth]{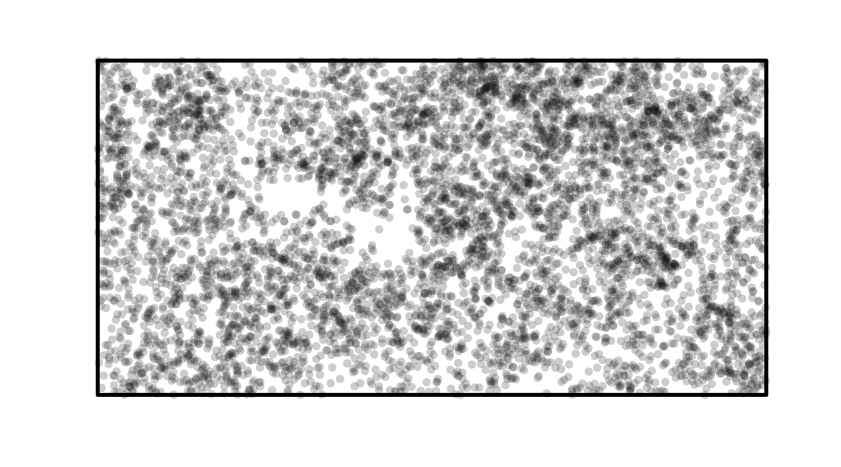} &	\includegraphics[width=0.33\textwidth]{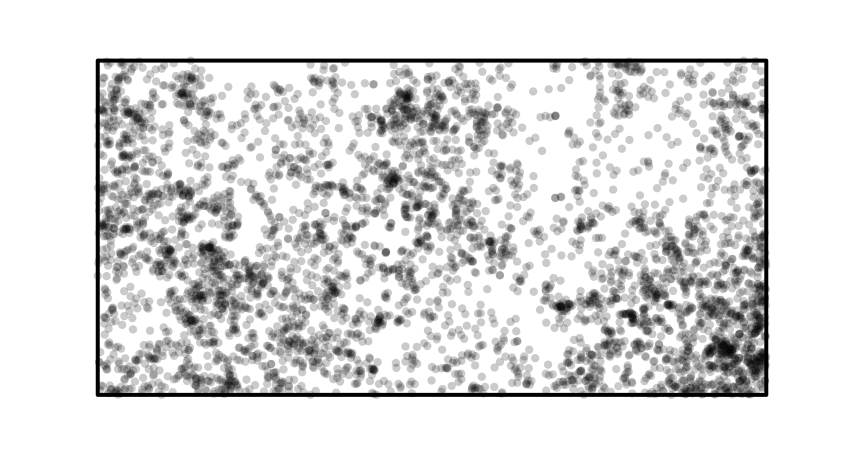} \\
		\end{tabular}
		\caption{Locations of 9 selected tree species (from left to right): 1st row: Capparis frondosa, Garcinia intermedia, and Hirtella triandra; 2nd row: Psychotria horizontalis, Protium tenuifolium, and Protium panamense; 3rd row: Mouriri myrtilloides, Swartzia simplex, and Tetragastris panamensis.}
		\label{fig:spec9}
	\end{figure*}

	Plots of 13 spatial covariates used for analysis are depicted in Figure~\ref{fig:cov}. Figure~\ref{fig:spec9} shows locations of the 9 selected tree species. 
	
\end{document}